# A Scattering Matrix Formalism to Model Periodic Heat Diffusion in Stratified Solid Media

**Tao Li and Zhen Chen***

Jiangsu Key Laboratory for Design & Manufacture or Micro/Nano Biomedical Instruments, School of Mechanical Engineering, Southeast University, Nanjing 210096, China

* correspondence should be addressed to: zhenchen@seu.edu.cn

**Abstract**

The transfer matrix formalism is widely used in modeling heat diffusion in layered structures. Due to an intrinsic numerical instability issue, which has not yet drawn enough attention to the heat transfer community, this formalism fails at high heating frequencies and/or in thick structures. Inspired by its success in modeling wave propagation, we develop a numerically-stable scattering matrix framework to model periodic heat diffusion in stratified solid media. As a concreate example, we apply this scattering matrix methodology to the $3\omega$ method. We first validate our framework using various well-known solutions. Next, we demonstrate the numerical stability of the framework using a configuration that resembles the three-dimensional stacked architecture for chip packing. Last, we propose synthetic "experiments" to exhibit, under certain circumstances, the merits of the scattering matrix formalism in extracting thermal properties.

## I. Introduction

The transfer matrix method has been widely applied to model periodic heating problems in layered structures, since Carslaw and Jaeger developed the analogous heat transfer formalism[1] to its electric[2] and elastic[3-4] counterparts. Feldman replaced the commonly used temperature and flux vector components with counter-propagating thermal waves[5], making it more analogous to the treatment in electromagnetices and optics. This matrix methodology has subsequently been implemented to analyze experimental data of well-known techniques for thermal measurements at the nanoscale, such as the $3\omega$ method[6-9], the time-domain thermoreflectance (TDTR) method[10-13] and the frequency-domain thermoreflectance (FDTR) method[14-15].

However, this transfer matrix formalism has an intrinsic numerical instability problem, which has not yet drawn enough attention to the heat transfer community. This problem arises from the fact that one diagonal element of the matrix exponentially decays while the other diagonal element exponentially blows up as the film thickness and/or the heating frequency increases. This issue has long been ignored since it only manifests itself when the heating frequency and/or the total thickness of the multilayer stack increases to some critical value that was of no practical interest. For example, for a $SiO_2$/Cu (50nm/50nm) alternating multilayer stack, the transfer matrix approach starts to break down only when the heating frequency increases to beyond GHz for microns thick stack (see, for example, Fig. 3c). This GHz heating frequency was thought to be unpractical before, but is becoming reasonable in the context of the ever-growing clock frequency of CPUs.

On the other hand, the wave community had long been suffered from this numerical instability problem[16-21]. Until 1988, did Ko and Inkson address this shortcoming by introducing the so-called scattering matrix scheme to model electron transmission in nonperiodic semiconductor heterostructures[22]. The essential idea is to regroup the inputs and outputs bridged by the matrix such that both of its diagonal elements exponentially decay, and thus avoid the numerical problem of the transfer matrix methodology. In 1999, Whittaker and Culshaw introduced this scattering matrix scheme to model the propagation of electromagnetic waves through multilayer structures[23]. However, to our knowledge, this numerically stable scattering matrix methodology has not yet been employed to treat the underlying heat transfer model of any experimental techniques.

Here we present the scattering matrix formalism to model periodic heat diffusion in stratified solid media. In Section II, we highlight the numerical instability problem of the transfer matrix scheme, and introduce the scattering matrix formalism using the one-dimensional diffusion equation. We consider the most general scenario in which the heater line is buried inside the multilayer structure. In Section III, we generalize the scattering matrix framework to multidimensions, with the 3 $\omega$ method as a concrete example. As sanity checks, we verify the framework using widely known solutions. In Section IV, we demonstrate the numerical stability of the scattering matrix formalism using a configuration that mimics the three-dimensional integration technology for processor architecture design. In Section V, we demonstrate with synthetic "experiments" that, in certain scenarios, this scattering matrix formalism could be employed to extract thermal properties, which is challenging for other schemes.

## II. Mathematical framework

In this section, we review the widely used transfer matrix method to solve heat diffusion problems [6-10], and point out its numerical instability issue. To address this shortcoming, we introduce the concept of the scattering matrix used in the electrical and the optical communities[22-23], and apply it to model heat diffusion in layered structures.

### A. Transfer matrix formalism and its numerical instability issue

We start with the one-dimensional (1D) heat diffusion equation,

$$\frac{1}{D}\frac{\partial T}{\partial t} = \frac{\partial^2 T}{\partial z^2}, \tag{1}$$

where $D(\equiv k/C)$ is the thermal diffusivity, $k$ is thermal conductivity, and $C$ is the volumetric specific heat.

According to the Linear Response Theorem[1], the temperature response of a linear system under a periodic heating flux ($q_{htr}e^{i\omega_H t}$) excitation reads as

$$\Delta T(z,t) = \theta(z)\, e^{i\omega_H t}, \tag{2}$$

where $\Delta T \equiv T - T_{ref.}$, and $T_{ref.}$ is a reference temperature, e.g. the ambient temperature, $T_\infty$. $\omega_H = 2\pi f_H = 4\pi f$ is the angular heating frequency, and $f$ is the frequency of the driving current.

Substituting Eq. 2 into Eq. 1, one obtains

$$\frac{i\omega_H}{D}\theta = \frac{\partial^2\theta}{\partial z^2}. \tag{3}$$

The general solution of Eq. 3 in layer $j$ of a multilayer structure (Fig. 1a) is

$$\theta_j(z) = a_j\Big|_{z=z_j} e^{-u_j(z-z_j)} + b_j\Big|_{z=z_j} e^{u_j(z-z_j)}, \tag{4}$$

where the coefficients, $a_j$ and $b_j$, depend on the reference point, here $z_j$. The magnitude of the complex wave vector

$$u_j = \sqrt{\frac{i\omega_H}{D_j}}, \tag{5}$$

is the inverse of the thermal penetration depth[24] in layer $j$.

Applying Fourier's law to Eq. 4, one obtains the corresponding heat flux, which, combining with Eq. 4, leads to[5]

$$\begin{bmatrix} a_{j+1}\Big|_{z=z_{j+1}} \\ b_{j+1}\Big|_{z=z_{j+1}} \end{bmatrix} = \mathbf{T_{j,j+1}} \times \begin{bmatrix} a_j\Big|_{z=z_j} \\ b_j\Big|_{z=z_j} \end{bmatrix}, \tag{6}$$

where the transfer matrix, $\mathbf{T_{j,j+1}} = \mathbf{E_{j,j+1}} \times \mathbf{\Phi_j}$, bridges the temperature and heat flux between layers $j$ and $j$ +1 (Fig. 1a), in which

$$\mathbf{E_{j,j+1}} = \frac{1}{2e_{j+1}} \begin{bmatrix} e_{j+1} + e_j & e_{j+1} - e_j \\ e_{j+1} - e_j & e_{j+1} + e_j \end{bmatrix} \tag{7}$$

is the effusivity matrix, and

$$\mathbf{\Phi_j} = \begin{bmatrix} e^{-u_j d_j} & 0 \\ 0 & e^{u_j d_j} \end{bmatrix} \tag{8}$$

is the phase matrix. Here $e_j = \sqrt{k_j C_j}$ is the effusivity, and $d_j$ is the thickness of layer $j$.

The phase matrix (Eq. 8) contains both exponentially decaying and growing entries, which leads to numerical instability issues when the heating frequency and/or the total thickness of the stratified medium increases to some critical values (see, for example, open circles in Fig. 3).

## B. Scattering matrix formalism

### 1. Basic idea

Motivated by the treatment of the propagation of electromagnetic waves in thick and lossy systems[23], here we develop a scattering matrix scheme to analyze periodic heating problems in multilayer structures. First, we rewrite Eq. 4 as

$$\theta_j(z) = a_j\big|_{z=z_j} e^{-u_j(z-z_j)} + b_j\big|_{z=z_j+d_j} e^{-u_j(z_j+d_j-z)}, \tag{9}$$

which chooses a different reference point, $z = z_j + d_j$, for $b_j$ (see Fig. 1b). As a result, the second term turns to exponentially decaying.

Second, we skew the input and output of Eq. 6 to arrive at

$$\begin{bmatrix} a_{j+1}\big|_{z=z_j+d_j} \\ b_j\big|_{z=z_j+d_j} \end{bmatrix} = \mathbf{S_{j,j+1}} \times \begin{bmatrix} a_j\big|_{z=z_j} \\ b_{j+1}\big|_{z=z_{j+1}+d_{j+1}} \end{bmatrix}, \tag{10}$$

where $\mathbf{S_{j,j+1}}$ is the so-called scattering matrix[22-23]. These two steps above ensure numerical stability of the scattering matrix formalism (see, for example, solid lines in Fig. 3).

## 2. Recursive relations to obtain $\mathbf{S_{i,j+1}}$ from $\mathbf{S_{i,j}}$

Analogous to the scattering matrix formalism in electromagnetics[23], we now develop the recursive relations to obtain $\mathbf{S_{1,N}}$ from $\mathbf{S_{1,1}}$ in heat transfer problems. By definition, $\mathbf{S_{1,1}}$ is an identity matrix. Therefore, the key here is to obtain $\mathbf{S_{i,j+1}}$ from $\mathbf{S_{i,j}}$.

First, for any layer *i* that is above layer *j* (Fig. 2), one applies Eq. 10 to obtain

$$\begin{bmatrix} a_j \\ b_i \end{bmatrix} = \mathbf{S_{i,j}} \times \begin{bmatrix} a_i \\ b_j \end{bmatrix}, \tag{11}$$

in which, for brevity, we neglect the reference points for the coefficients, $a_i, b_i$ $a_j$, and $b_j$.

Second, combining the temperature (Eq. 9) and its corresponding flux in layer *j*, one obtains

$$\begin{bmatrix} \theta_j(z) \\ q_j(z) \end{bmatrix} = \mathbf{M_j} \times \begin{bmatrix} a_j e^{-u_j(z-z_j)} \\ b_j e^{-u_j(z_j+d_j-z)} \end{bmatrix}, \tag{12}$$

where

$$\mathbf{M_j} = \begin{bmatrix} 1 & 1 \\ \gamma_j & -\gamma_j \end{bmatrix}, \tag{13}$$

and

$$\gamma_j = k_{j,z} u_j \tag{14}$$

Third, applying Eq. 12 to layer *j*+1, i.e., replacing the index *j* with *j*+1, and assuming continuity condition for both the temperature and flux at the interface between layers *j* and *j*+1, one obtains

$$\begin{bmatrix} a_j e^{-u_j d_j} \\ b_j \end{bmatrix} = \mathbf{I_{j,j+1}} \times \begin{bmatrix} a_{j+1} \\ b_{j+1} e^{-u_{j+1} d_{j+1}} \end{bmatrix}, \tag{15}$$

where

$$\mathbf{I_{j,j+1}} = \mathbf{M_j^{-1}} \times \mathbf{M_{j+1}} \tag{16}$$

is the interface matrix. To incorporate the thermal contact resistance, $R''_{c,j-j+1}$ (with SI units of $[\mathrm{m^2KW^{-1}}]$), between layers $j$ and $j+1$, one updates Eq. 16 to

$$\mathbf{I'_{j+1,j}} = \mathbf{M_j^{-1}} \times \mathbf{M'_{j+1}}, \tag{17}$$

where

$$\mathbf{M'_{j+1}} = \begin{bmatrix} 1 + R''_{c,j-j+1}\gamma_{j+1} & 1 - R''_{c,j-j+1}\gamma_{j+1} \\ \gamma_{j+1} & -\gamma_{j+1} \end{bmatrix}. \tag{18}$$

At last, combining Eqs. 11 and 15, one develops the recursive relations to obtain $\mathbf{S_{i,j+1}}$ from $\mathbf{S_{i,j}}$, as derived in Appendix B and summarized in Table I.

## 3. The driving periodic heating flux and boundary conditions

We consider a general case, in which the driving flux ($q_{htr}e^{i\omega_H t}$) is exerted at the interface between layers $j$ and $j+1$ by a heater with negligible thickness (Fig. 2). In this case, Eq. 15 is updated to

$$\mathbf{M_{j+1}} \times \begin{bmatrix} a_{j+1} \\ b_{j+1}e^{-u_{j+1}d_{j+1}} \end{bmatrix} - \mathbf{M_j} \times \begin{bmatrix} a_j e^{-u_j d_j} \\ b_j \end{bmatrix} = q_{htr}\begin{bmatrix} 0 \\ 1 \end{bmatrix}. \tag{19}$$

We discuss two types of commonly used boundary conditions, an adiabatic boundary condition and an isothermal boundary condition. In the latter we assume the top and bottom boundaries are fixed at the same $T_{ref.}$, which results in $\theta_{bdy} = 0$.

For the top boundary, one has

$$a_1 - c_{bdy}b_1e^{-u_1d_1} = 0. \tag{20}$$

Likewise, for the bottom boundary one arrives at

$$c_{bdy}a_Ne^{-u_Nd_N} - b_N = 0. \tag{21}$$

where $c_{bdy} = 1$ or -1 for an adiabatic or isothermal boundary condition.

If the top or bottom layer is semi-infinite, Eq. 20 or 21 reduces to

$$a_1 = 0 \text{ or } b_N = 0, \tag{22}$$

which is equivalent to set $c_{bdy} = 0$.

We summarize different values of $c_{bdy}$ according to different boundary conditions,

$$c_{bdy} = \begin{cases} 1, & \text{adiabatic} \\ -1, & \text{isothermal.} \\ 0, & \text{semi}-\text{infinite} \end{cases} \tag{23}$$

## 4. Recipe: solving the temperature field

Using the recursive relations (Table I), the driving flux (Eq. 19), and the boundary conditions (Eqs. 20-23), one can solve the temperature profile of the multilayer structure, following the recipe below.

First, using Eq. 11, one obtains

$$\begin{bmatrix} a_j \\ b_1 \end{bmatrix} = \mathbf{S_{1,j}} \times \begin{bmatrix} a_1 \\ b_j \end{bmatrix}, \tag{24}$$

and

$$\begin{bmatrix} a_N \\ b_{j+1} \end{bmatrix} = \mathbf{S_{j+1,N}} \times \begin{bmatrix} a_{j+1} \\ b_N \end{bmatrix}, \tag{25}$$

where $\mathbf{S_{1,j}}$ and $\mathbf{S_{j+1,N}}$ can be obtained using the recursion relations in Table I.

Next, substituting the boundary conditions (Eqs. 20 and 21) into Eqs.24 and 25, one obtains

$$a_j = \alpha b_j, \tag{26}$$

$$b_{j+1} = \beta a_{j+1}, \tag{27}$$

where

$$\alpha = \mathbf{S_{1,j}}(1,2) + c_{bdy} \frac{\mathbf{S_{1,j}}(1,1)\mathbf{S_{1,j}}(2,2)e^{-u_1 d_1}}{1 - c_{bdy}\mathbf{S_{1,j}}(2,1)e^{-u_1 d_1}}, \tag{28}$$

$$\beta = \mathbf{S_{j+1,N}}(2,1) + c_{bdy} \frac{\mathbf{S_{j+1,N}}(1,1)\mathbf{S_{j+1,N}}(2,2)e^{-u_N d_N}}{1 - c_{bdy}\mathbf{S_{j+1,N}}(1,2)e^{-u_N d_N}}. \tag{29}$$

Third, combining Eqs. 26-29, and 19, one solves $a_j$, $b_j$, $a_{j+1}$, and $b_{j+1}$ (Table II).

At last, Substituting $a_j$, $b_j$, $a_{j+1}$, and $b_{j+1}$ back to Eqs. 24 and 25, one obtains $a_1$, $b_1$, $a_N$, and $b_N$. The coefficients of all the other layers can be obtained from Eq. 24 or 25 by replacing the index $j$ or $j+1$ to other indices of interest. Knowing these coefficients, one obtains the temperature field of the multilayer structure (Eq. 9).

## III. Apply the framework to the 3 $\omega$ method

In section II, we develop the scatter matrix formalism for the one-dimensional (1D) diffusion equation. We now generalize the framework to two-dimensional (2D) in this section, using the 3 $\omega$ method as a concrete example.

### A. From 1D to 2D

Although the original $3\omega$ method was derived in a cylindrical coordinate[1, 24], here, to take into account the anisotropy of films and substrates, we follow Ref. [25] and choose a Cartesian coordinate. For simplicity, we assume the principal axes of the thermal conductivity tensor are aligned with the natural cartesian coordinate system defined by the heater line and sample surface, i.e., the thermal conductivity tensor is diagonalized. We briefly comment on the more general case of an anisotropic thermal conductivity tensor with finite off-diagonal terms in Appendix C.

Isolating the time dependence and performing the Fourier cosine transform (along the $x$-axis, see Fig. 1) to the 2D diffusion equation, one obtains

$$\frac{i\omega_H}{D_z}\tilde{\theta} = -n_{xz}\lambda^2\tilde{\theta} + \frac{\partial^2\tilde{\theta}}{\partial z^2}, \tag{30}$$

where $n_{xz} = k_x/k_z$ is the anisotropy ratio, and $k_x$ and $k_z$ are the in-plane and the cross-plane thermal conductivities. As compared to Eqs. 3 and 5, Eq. 30 is now an ordinary differential equation with an updated complex wave vector[7]

$$\tilde{u}_j = \sqrt{n_{j,xz}\lambda^2 + \frac{i\omega_H}{D_{j,z}}}, \tag{31}$$

where $D_{j,z} = \frac{k_{j,z}}{C_j}$ is the thermal diffusivity of layer *j* along *z*-axis.

For the general case where the heater is located at the interface between layers *j* and *j*+1 (Fig. 2), one can still apply Eq. 9 to obtain the spatial component of the temperature of the heater in Fourier space:

$$\tilde{\theta}_{htr} = \tilde{\theta}_{j+1}\big|_{z=z_{j+1}} = a_{j+1} + b_{j+1}e^{-\tilde{u}_{j+1}d_{j+1}}$$

$$= \left(a'_{j+1} + b'_{j+1}e^{-\tilde{u}_{j+1}d_{j+1}}\right)\tilde{q}_{htr}, \tag{32}$$

where the normalized coefficients, $a'_{j+1} = a_{j+1}/\tilde{q}_{htr}$ and $b'_{j+1} = b_{j+1}/\tilde{q}_{htr}$, are computed using Table II. As compared to the 1D scenario, here the only difference is to replace $u$ (Eq. 5) with $\tilde{u}$ (Eq. 31), and to replace $q_{htr}$ (Table II) with its Fourier cosine transform, $\tilde{q}_{htr}$, which will be illustrated with examples in Section III-C.

Transforming Eq. 32 back to the real space, one obtains the spatial component of the temperature of the heater:

$$\theta_{htr}(x) = \frac{2}{\pi}\int_0^\infty \tilde{\theta}_{htr}\cos(\lambda x)\, d\lambda. \tag{33}$$

### B. Most common scenario: line heater on sample

Section II-B gives a recipe to implement the scattering matrix formalism, in which the heater is embedded inside the sample (Fig. 2). To apply this general framework to a special and also most common scenario, line heater on sample, here we highlight some key steps to illustrate how to use Tables I-II.

First, treat the air/vacuum above the sample as a semi-infinite layer[5], and denote it as layer 0. Since the effusivity $e_{air} \ll e_{solid}$, we neglect the heat conduction through layer 0, and thus using Eq. 13 we obtain

$$\mathbf{M_0} = \begin{bmatrix} 1 & 1 \\ 0 & 0 \end{bmatrix}. \tag{34}$$

Next, replace, in Eq. 28 and Eq. 29, both the subscripts 1 and *j* with 0 in matrices $\mathbf{S_{1,j}}$ and $\mathbf{S_{j+1,N}}$, the complex wave vector $u_j$ (or, $\tilde{u}_j$ in the 2D scenario here), and the layer thickness $d_j$. Therefore one obtains $\alpha = 0$ (Eq. 28), regardless of the boundary condition. On the other hand, $\beta$ (Eq. 29) depends on the boundary condition and $\mathbf{S_{1,N}}$ that can be obtained from the recursion relations in Table I.

At last, solving for $a_1$ and $b_1$ using Table II, and substituting them into Eqs. 32 and 33, one obtains

$$\theta_{htr-on-top}(x) = \frac{2}{\pi}\int_0^\infty \frac{1+\beta e^{-\tilde{u}_1 d_1}}{\tilde{\gamma}_1\left(1-\beta e^{-\tilde{u}_1 d_1}\right)}\tilde{q}_{htr}\cos(\lambda x)\, d\lambda, \tag{35}$$

where $\tilde{\gamma}_1 = k_1\tilde{u}_1$ is defined in Eq. 14.

## C. Verification of the framework

We verify the framework using widely known solutions, as discussed in detail in the following and summarized in Table III. Note here we treat all the substrates to be semi-infinite.

We consider two types of heaters. First, if the line heater is infinitely-narrow, one has

$$q_{htr} = \frac{P_0}{l}\delta(x), \tag{36}$$

where $P_0$ is the root-mean-square value of the Joule heating power (with SI units of [W]). Performing the Fourier cosine transform to Eq. 36, one obtains

$$\tilde{q}_{htr} = \int_0^\infty \frac{P_0}{l}\delta(x)\cos(\lambda x)\,dx = \frac{P_0}{2l}. \tag{37}$$

Second, if the line heater has a finite width, $w$, one has

$$q_{htr} = \begin{cases} \frac{P_0}{wl}, & |x| \le \frac{w}{2} \\ 0, & |x| > \frac{w}{2}, \end{cases} \tag{38}$$

and its corresponding Fourier cosine transform

$$\tilde{q}_{htr} = \int_0^\infty \frac{P_0}{wl}\cos(\lambda x)\,dx = \frac{P_0}{2l}\frac{\sin\left(\lambda\frac{w}{2}\right)}{\left(\lambda\frac{w}{2}\right)}. \tag{39}$$

### 1. Heater on semi-infinite substrate (1st row of Table III)

In this scenario, one has $\mathbf{S_{j+1,N}} = \mathbf{S_{1,1}} = \mathbf{I}$, and thus $\beta = \mathbf{S_{1,1}}(2,1) = 0$ for a semi-infinite substrate (Eq. 29). For an infinitely-narrow heater, substituting Eq. 37 and $\beta = 0$ into Eq. 35, one obtains[26]

$$\theta_{htr-on-top}(x) = \frac{P_0}{\pi l\sqrt{k_x k_z}} K_0\left(\frac{1}{\sqrt{n_{xz}}}\sqrt{\frac{i\omega_H}{D_z}}x\right), \tag{40}$$

where $K_0$ is the modified Bessel function of the second kind. For an isotropic substrate, $k_x = k_z = k$ and thus $n_{xz} \equiv k_x/k_{,z} = 1$, Eq. 40 reduces to

$$\theta_{htr-on-top}(x) = \frac{P_0}{\pi lk} K_0\left(\sqrt{\frac{i\omega_H}{D}}x\right), \tag{41}$$

which recovers the well-known solution of the classical $3\omega$ method[1, 24].

Likewise, for a heater with finite width ($w$), substituting Eq. 38 and $\beta = 0$ into Eq. 35, and averaging across the heater width, one obtains

$$\theta_{htr-on-top,avg.} = \frac{P_0}{\pi l\sqrt{k_x k_z}} \int_0^\infty \frac{1}{\sqrt{\lambda^2 + \frac{1}{n_{xz}}\frac{i\omega_H}{D_z}}} \frac{\sin^2\left(\lambda\frac{w}{2}\right)}{\left(\lambda\frac{w}{2}\right)^2} d\lambda, \tag{42}$$

and its corresponding isotropic form,

$$\theta_{htr-on-top,avg.} = \frac{P_0}{\pi lk} \int_0^\infty \frac{1}{\sqrt{\lambda^2 + \frac{i\omega_H}{D}}} \frac{\sin^2\left(\lambda\frac{w}{2}\right)}{\left(\lambda\frac{w}{2}\right)^2} d\lambda, \tag{43}$$

which recovers Eq. 8 of Ref. [24]. If the thermal penetration depth,

$$L_{penetr.} = \sqrt{D/\omega_H}, \quad (44)$$

is much larger than the half width of the heater, i.e. $L_{penetr.} \gg w/2$, Eq. 43 could be further simplified to

$$\theta_{htr-on-top,avg.} \approx \frac{P_0}{\pi l k}\left[\frac{1}{2}ln\left(\frac{D}{(w/2)^2}\right) + \eta - \frac{1}{2}ln(\omega_H) - i\frac{\pi}{4}\right], \quad (45)$$

which recovers Eq. 15 of Ref. [26] or Eq. 1 of Ref. [27] .

**2. Heater on film-substrate stack (2nd row of Table III)**

Here we consider a line heater with finite width, $w$. In this scenario, one has $\mathbf{S_{j+1,N}} = \mathbf{S_{1,2}}$, and thus $\beta = \mathbf{S_{1,2}}(2,1)$ for a semi-infinite substrate (Eq. 29). Using the recursion relations in Table I, one obtains $\mathbf{S_{1,2}}(2,1) = \frac{\tilde{\gamma}_1-\tilde{\gamma}_2}{\tilde{\gamma}_1+\tilde{\gamma}_2}e^{-\tilde{u}_1 d_1}$.

Following the same procedure above, one recovers the solution using the transfer matrix formalism (Eq. 1 of Ref. [6]),

$$\theta_{htr-on-top,avg.} = \frac{P_0}{\pi l}\int_0^\infty \frac{1}{\gamma_1}\frac{B_{12}^+ + B_{12}^-}{B_{12}^+ - B_{12}^-}\frac{\sin^2\left(\lambda\frac{w}{2}\right)}{\left(\lambda\frac{w}{2}\right)^2}d\lambda, \quad (46)$$

where

$$B_{12}^{\pm} = \frac{1}{2\tilde{\gamma}_1}\left(\tilde{\gamma}_1 \pm \tilde{\gamma}_2 + R''_{c,1-2}\tilde{\gamma}_1\tilde{\gamma}_2\right)e^{\pm\tilde{u}_1 d_2}, \quad (47)$$

with 1 denoting the film, and 2 denoting the substrate (see the schematic in Table III). Note here we include the thermal contact resistance between the film and the substrate, $R''_{c,film-sub.}$.

Equation 46 can be regrouped to

$$\theta_{htr-on-top,avg.} = \theta_{sub.} + \theta_{film}, \quad (48)$$

where $\theta_{sub.}$ is computed using Eq. 42, and

$$\theta_{film} = \theta_{film,1D} \times s, \quad (49)$$

where the dimensionless scaling factor, $s$ (see Appendix E), indicates how much deviation the 1D approximation,

$$\theta_{film,1D} = P_0 d_{film}/(wlk_{film,z}) + P_0 R''_{c,film-sub.}/(wl), \quad (50)$$

is away from the true $\theta_{film}$. As compared to previous works[7, 25], our analysis here is not restricted to the zero-$f_H$ assumption.

With further approximations (see Appendix E), and assuming isotropic properties of the sample, one reduces Eq. 48 to the well-known series-resistor solution

$$\theta_{htr-on-top,avg.} \approx \frac{P_0}{\pi l k_{sub.}}\left[\frac{1}{2}ln\left(\frac{D_2}{(w/2)^2}\right) + \eta - \frac{1}{2}ln(\omega_H) - i\frac{\pi}{4}\right] + \frac{P_0}{wl}\frac{d_{film}}{k_{film}} + \frac{P_0}{wl}R''_{c,film-sub.}. \quad (51)$$

Note here as compared to Eqs. 1 and 2 of Ref. [27], Eq. 51 includes the effect of the thermal contact resistance, $R''_{c,film-sub.}$.

### 3. Line heater sandwiched by two film-substrate stacks (3rd row of Table III)

This configuration has to be solved using the general framework (Eqs. 32 and 33, instead of Eq. 35). Here we again consider a line heater with finite width, $w$. Using Tables I-II, we compute the **S**-matrices, and correspondingly $\alpha$ and $\beta$, and finally $a'_{j+1}$ and $b'_{j+1}$ (Table D1).

Substituting $a'_{j+1}$, $b'_{j+1}$, into Eqs. 32 and 33, and performing an average over the width of the heater, one obtains

$$\theta_{htr-sandwiched,avg.} = \frac{P_0}{\pi l}\int_0^\infty \frac{1}{\tilde{\gamma}_2\frac{(B_{21}^+-B_{21}^-)}{(B_{21}^-+B_{21}^+)}+\tilde{\gamma}_3\frac{(B_{34}^+-B_{34}^-)}{(B_{34}^++B_{34}^-)}}\frac{\sin^2\left(\lambda\frac{w}{2}\right)}{\left(\lambda\frac{w}{2}\right)^2}d\lambda. \quad (52)$$

where $B_{21}^{\pm}$ and $B_{34}^{\pm}$ are evaluated using Eq. 47 with updated subscripts. Note here 1, 2, 3, and 4 denote the upper substrate, the upper film, the lower film, and the lower substrate, respectively (see the schematic in Table III). After some re-arrangements, Eq. 52 recovers the solution using the transfer matrix formalism[8-9].

Note here for simplicity we ignore the contact thermal resistance between layers. If the upper substrate (layer 1) and film (layer 2) are the same materials, i.e. $k_1 = k_2 = k_{upper}$ and $D_1 = D_2 = D_{upper}$, and so are the lower film (layer 3) and substrate (layer 4), i.e. $k_3 = k_4 = k_{lower}$ and $D_3 = D_4 = D_{lower}$, Eq. 52 reduces to $\theta_{htr-sandwiched,avg.} =$

$$\frac{P_0}{\pi l}\int_0^\infty \frac{1}{\sqrt{k_{upper,x}k_{upper,z}}\sqrt{\lambda^2+\frac{i\omega_H}{n_{upper,xz}D_{upper,z}}}+\sqrt{k_{lower,x}k_{lower,z}}\sqrt{\lambda^2+\frac{i\omega_H}{n_{lower,xz}D_{lower,z}}}}\frac{sin^2\left(\lambda\frac{w}{2}\right)}{\left(\lambda\frac{w}{2}\right)^2}d\lambda. \quad (53)$$

For the special case of $n_{upper,xz}D_{upper,z} = n_{lower,xz}D_{lower,z} = n_{xz}D_z$, also known as the boundary mismatch approximation (BMA)[28], one simplifies Eq. 53 to

$$\theta_{htr-sandwiched,avg.} = \frac{P_0}{\pi l\left(\sqrt{k_{upper,x}k_{upper,z}}+\sqrt{k_{lower,x}k_{lower,z}}\right)}\int_0^\infty \frac{1}{\sqrt{\lambda^2+\frac{i\omega_H}{n_{xz}D_z}}}\frac{sin^2\left(\lambda\frac{w}{2}\right)}{\left(\lambda\frac{w}{2}\right)^2}d\lambda. \quad (54)$$

If one further assumes isotropic properties, Eq. 54 reduces to

$$\theta_{htr-sandwiched,avg.} = \frac{P_0}{\pi l(k_{upper}+k_{lower})}\int_0^\infty \frac{1}{\sqrt{\lambda^2+\frac{i\omega_H}{D_z}}}\frac{sin^2\left(\lambda\frac{w}{2}\right)}{\left(\lambda\frac{w}{2}\right)^2}d\lambda, \quad (55)$$

which recovers Eq. B2 of Ref. [28].

## IV. Demonstration of the numerical stability of the framework

To demonstrate the numerical stability of the scattering matrix scheme, here we consider a 20 $\mu m$-wide line heater sandwiched by two identical superlattice-on-substrate structures (Fig. 3a). We choose a 500 $\mu m$-thick silicon substrate, and 50nm/50nm-thick $SiO_2$/Cu alternating films as the building blocks of the superlattice, with corresponding thermal conductivities[29] of $k_{Si}$=148 W/m-K, $k_{SiO2}$=1.38 W/m-K, and $k_{Cu}$= 401 W/m-K, and the thermal contact resistance[30] of $R^{"}_{c,SiO_2-Cu} = 1 \times 10^{-8}$ $m^2$ /W-K. Note here we neglect the possible size effect on $k_{Cu}$ for the nanoscale Cu film. This configuration mimics the three-dimensional stacked architecture for chip packing[31-32].

Figure 3b shows the temperature rise of the heater normalized by the corresponding periodic heating power, i.e. the thermal resistance, $R$ $(= \theta_{htr,avg.}/P_0)$, as a function of the number of periods $(N = N_1 + N_2)$ at a fixed heating frequency, $f_H$ = 10 Hz, calculated using both the transfer matrix scheme (open circles) and the scattering matrix scheme (solid line). While both approaches work well and agree with each other for configurations with small $N$, the transfer matrix scheme breaks down at a critical total number of periods, $N_c$ = 44. In contrast, the scattering matrix scheme works reliably regardless of $N$, as expected. We note that although the exact value of $N_c$ may depend on the specific numerical method to implement the integration (Eq. 33) and/or the specification of the computer, the transfer matrix approach is destined to fail at some point when $N$ keeps increasing, as discussed in Section II-A.

In addition to the failure for thick structures, the transfer matrix scheme is also destined to break down at high heating frequencies. This is confirmed by Fig. 3c, in which we fix $N_1 = N_2 = 20$ (and thus $N = 40$) and plot the $R$ vs. $f_H$ relation. While the transfer matrix approach (open circles) fails beyond a critical heating frequency, $f_{H,c} \approx$ 3GHz, the scattering matrix approach (solid line) is robust over the entire frequency range of interest. Like $N_c$ in Fig. 3b, this $f_{H,c}$ is destined to appear somewhere, although its exact value may depend on the specific integration method and/or the specification of the computer. Generalizing the results of Figs. 3b-c leads to the regime map in Fig. 3d. In the lower regime (white), both the transfer matrix approach and the scattering matrix approach work well in computing the temperature response of the heater, while in the other regime (gray shaded), the transfer matrix formalism breaks down but the scattering matrix formalism is still robust.

We briefly comment on the periodic heating in the GHz regime and beyond. First, this high frequency heating is becoming realistic in integrated circuits in the context that the clock frequency of CPU has already been in the same order of magnitude. Second, the thermal wave in this regime is well localized inside the top layer for $f_H > f_{H,c}$. For example, the critical penetration depth, $L_{penetr.,c} = \sqrt{D_{Si}/2\pi f_{H,c}} \approx$ 6.6nm (Eq. 44), which is much smaller than the thickness of the upper and lower SiO2 layers adjacent to the heater, $d_{SiO_2}$ =50 nm. Correspondingly, an approximate solution is to treat the $SiO_2$ layers as semi-infinite substrates and compute the temperature rise of the heater using Eq. 43. Third, although the ballistic phonon transport should be considered in this regime[33-36]. it is still the paradigm of experimental techniques in which solutions of the diffusion equation are used to fit for effective thermal properties from experimental data[37-39].

## V. Synthetic "experiments" to extract thermal properties

In addition to the application to the complicated multilayer structure in Fig. 3, the scattering matrix formalism can manifest its merits in experimentally extracting thermal properties even for the simple film-on-substrate configuration under certain circumstances. To demonstrate these merits, we generate three sets of synthetic "experimental" data points using a 3D finite element model (FEM)[40], including $\pm 1\%$ of "experimental noise" (open circles in Fig. 4). In the 3D FEM model, we fix the thickness of the silicon substrate to be $d_{Si} = 500\mu m$, and set the width and the length (into the page) of the heater to be $w = 20\mu m$ and $l = 2{,}000\mu m$, respectively. We exert a driving periodic heat flux on the top surface of the heater, and an isothermal boundary condition ($T_\infty$) on the bottom surface of the silicon substrate, with all the other surfaces insulated. The input thermal properties to the FEM model are summarized in Table IV. We verify that the size of the film and the substrate is large enough to ensure numerical convergence. We choose a heating frequency range of 500-2,000 Hz to ensure the assumption of a narrow heater on a semi-infinite substrate (Appendix F).

For all three scenarios (rows 1-3 of Fig. 4), the thermal conductivity of the silicon substrate, $k_{Si}$, is obtained by the slope method[24], in which the slope of the linear relation of $R_{synth.-expt.}$ vs. $\ln(f_H)$ gives $k_{Si}$. Here $R_{synth.-expt.} = \theta_{htr,avg.}/P_0$, and $\theta_{htr,avg.}$ is the amplitude of the temperature oscillation averaged across the width of the heater line. To extract $k_{film}$, we apply three approaches. In the first approach (1st column of Fig. 4), we compute the thermal resistance of the substrate, $R_{Si-model}$, using Eq. 45 and $k_{Si}$ from the slope method. Then we obtain the thermal resistance of the thin film, $R_{film} = R_{synth,.-expt.} - R_{Si-model}$, and correspondingly $k_{film}$ assuming 1D heat conduction across the thin film[27, 41]. Note in this 1D approximation approach, $R_{film}$ and thus $k_{film}$ include the effect of the thermal contact resistance between the film and the substrate, $R''_{c,SiO_2-Si}$. In the second (2nd column) or the third (3rd column) approach, we treat $k_{film}$ as the only free parameter to fit Eq. 46 (or Eq. 35) to the synthetic "experiments" using the transfer matrix (or the scattering matrix) formalism, respectively.

The first scenario (upper row of Fig. 4) shows a sanity check using a standard configuration of the classic $3\omega$ measurement, a $0.3\mu m$-thick low-$k$ $SiO_2$ thin film on a high-$k$ silicon substrate. In this scenario, all the three methods succeed in extracting $k_{SiO_2}$. Note in this scenario we include a thermal contact resistance, $R''_{c,SiO_2-Si} = 1 \times 10^{-8}$m$^2$-K/W[27], in the synthetic "experiment" (i.e. the FEM model; Table IV), but neglect it in the fitting scheme using either the transfer matrix or the scattering matrix formalism, since the intrinsic $k_{SiO_2}$ dominates in this scenario.

The second scenario (middle row of Fig. 4) replaces the $SiO_2$ film with a $4\,\mu m$-thick high-$k$ diamond film, which closely resemble a proposed configuration of post-silicon heat sinks[42-43]. It is very challenging to measure this high-$k$-film-on-substrate configuration, because the $3\omega$ method gradually loses its sensitivity to $k_{film}$ as it surpasses $k_{Si}$ (Appendix G). In addition, the anisotropic nature of $k_{Diamond}$ and the exist of $R^{''}_{c,Diamond-Si}$ that cannot be neglected in this scenario because of the high $k_{Diamond}$ make the measurement even more complicated. In the following, we will show that the scattering matrix framework offers a way to extract the cross-plane thermal conductivity of the diamond film, $k_{Diamond,z}$, for this high-$k$-film-on-substrate configuration, regardless of the thickness of the diamond film.

As can be seen from Fig. 4d, the 1D approximation approach fails in measuring this high-$k$-film-on-substrate configuration, because $R_{synth,.-expt.} < R_{Si-model}$, which implies a negative $R_{film}$! We now come to a dilemma: even if we neglect $R^{''}_{c,Diamond-Si}$, how could the effective resistance, $R_{synth,.-expt.} = R_{Si-model} + R_{film}$, be smaller than $R_{Si-model}$? We qualitatively resolve this seemingly paradoxical problem using the following two-step argument. First, replace the $SiO_2$ film in the first scenario (1st row of Fig. 4) with a silicon film. If one neglects the size effect of $k_{Si-film}$ (i.e. we assign $k_{Si-film} = k_{Si-substrate}$ here), it is not hard to conclude that $R_{synth,.-expt.} = R_{Si-model}$ in the heating frequency range that ensures $L_{penetr.} \ll d_{Si}$, because the combination of a silicon thin film and a semi-infinite silicon substrate is no more than a single semi-infinite silicon substrate. Second, replace the silicon film with a diamond film. Now it is easy to conclude that $R_{synth,.-expt.} < R_{Si-model}$, since $k_{Diamond,z} > k_{Si}$. Therefore, in this high-$k$-film-on-substrate scenario, the 1D approximation approach fails. We will return to this interpretation quantitatively in the context of Fig. 5 shortly.

We now apply the transfer matrix and the scattering matrix methods to fit for $k_{Diamond,z}$. Since $k_{Si}$ and $C_{Si}$ had been accurately measured and well documented, the sensitivity analysis in Appendix G indicates that, among the remaining input parameters, $k_{Diamond,z}$ is most sensitive to $R^{''}_{c,Diamond-Si}$ and $k_{Diamond,x}$. We obtain $R^{''}_{c,Diamond-Si}$ by applying the 1D approximation to a $0.1\mu m$-thick diamond film on silicon substrate configuration, in which $R_{film}$ is dominated by $R^{''}_{c,Diamond-Si}$ (see Appendix G for more details). We further assume $k_{Diamond,x}$ is obtained using the suspended microfabricated-device method with typical measurement uncertainty of 10%[44]. Fitting either the transfer matrix or the scattering matrix model to one set of typical synthetic "experiment", one obtains the only fitting parameter, $k_{Diamond,z}$, with 9.4% error as compared to the input value (Figs. 4e and 4f; Table IV).

The last scenario (lower row of Fig. 4) increases the thickness of the diamond film, $d_{film}$, from $4\mu m$ to $7.5\mu m$. In this case, the "transfer matrix" approach also fails due to its intrinsic numerical instability, as discussed in Section II-A. On the other hand, the scattering matrix approach is still robust in fitting $k_{Diamond,z}$ with an error of 2.7% for one set of typical synthetic "experiment" (Figs. 4i; Table IV). In Appendix G, we conduct multiple synthetic "experiments" to analyze the confidence interval of these "measurements".

At last, we turn to quantifying our qualitative interpretation to the dilemma in Figs. 4d and 4g. Recall Eq. 49, the full expression of the temperature rise of the film, $\theta_{film}$. Figure 5a shows a contour plot of the dimensionless scaling factor $s$ (Eq. E1 of Appendix E), which indicates how much deviation the 1D approximation, $\theta_{film,1D}$, is away from the true $\theta_{film}$. The *x*-axis is the contrast factor between the thermal conductivity of the film and the substrate, $\Pi_1 = k_{film,z}/k_{sub.,z}$, and the *y*-axis is the spreading factor of the film, $\Pi_4 = \sqrt{k_{film,x}/k_{film,z}}\frac{d_{film}}{w/2}$. Here we choose $f_H =$ 500Hz, leading to $\Pi_2 = d_{film}/L_{penetr.,film} = 0.011$, $\Pi_3 = d_{film}/L_{penetr.,sub.} = 0.024$, $\Pi_5 = \sqrt{k_{sub.,x}/k_{sub.,z}}\frac{d_{film}}{w/2} = 0.40$ and $\Pi_6 = R''_{c,film-sub.}k_{film}/d_{film} = 2.1$, according to the dimensions of the heater and the sample (Fig. 4) and the thermal properties summarized in Table IV. The fact that $\Pi_3 \ll 1$, $\Pi_5 < 1$, and $L_{penetr.,sub.} \ll d_{sub.}$ may suggest that 1D approximation is justified. However, Fig. 5a shows that the series-resistor solution (Eq. 51) could still introduce significant errors when the contrast factor ($\Pi_1$) and/or the spreading factor of the film ($\Pi_4$) are relatively large. This result confirms the negative $R_{film}$ observed in Figs. 4d, which is indicated using a triangle in Fig. 5a. Likewise, Fig. 5b shows a contour plot of $s$ with the *y*-axis replaced by the dimensionless thermal contact resistance, $\Pi_6 = R''_{c,film,sub.}k_{film}/d_{film}$. With $\Pi_2 = 0.021$, $\Pi_3 = 0.045$, $\Pi_4 = 0.34$, and $\Pi_5 = 0.75$, this contour plot confirms the negative $R_{film}$ observed in Figs. 4g, which is indicated using a circle in Fig. 5b.

## VI. Conclusions and Discussion

To circumvent the intrinsic numerical instability problem of the transfer matrix formalism, we have generalized the scattering matrix methodology of wave propagation to model periodic heat diffusion in layered structures. We applied this scattering matrix framework to the $3\omega$ method, and validated it with various well-known solutions. Then we demonstrated its numerical stability in modeling the periodic heating of a thick 3D stacked architecture for chip packing at high heating frequency. We also proposed synthetic "experiments" to show that this framework can be used to extract thermal properties in certain circumstances when the conventional techniques break down.

We conclude by commenting on the potential application of this scattering matrix framework to two other popular experimental techniques, the time-domain thermoreflectance (TDTR)[10-13] and the frequency-domain thermoreflectance (FDTR)[14-15]. As compared to the experimental setup of the $3\omega$ method, TDTR and FDTR can be pushed to much higher heating frequencies, typically at tens of to hundreds of MHz[37, 39]. Moreover, TDTR and FDTR rely more on the transfer matrix formalism to fit for thermal properties of multilayer structures[10-15, 39]. These facts make the scattering matrix formalism an ideal choice for these laser techniques to avoid numerical instability issues of the transfer matrix methodology. To summarize the analogy and the contrast, we briefly outline the key steps of implementing the scattering matrix framework to the 1D planar heating, the 2D 3 $\omega$ method, and the 3D TDTR/FDTR in Table V. For the 3D case (last column of Table V), the Fourier cosine transform is replaced with a Hankel transform, and the complex wave vector, $\tilde{u}$, is updated correspondingly[10-15].

**Acknowledgements**

This work was supported in part by National Natural Science Foundation of China (51776038).

**Tables and Figures**

TABLE I. Recursive relations to obtain $\mathbf{S_{i,j+1}}$ from $\mathbf{S_{i,j}}$ (see detailed derivation in Appendix B. By definition, $\mathbf{S_{i,i}}$ is an identity matrix. The interface matrix, $\mathbf{I_{j,j+1}}$, is defined in Eq. 16 (or Eq. 17 if considering contact resistance), and the complex wave vector, $u_j$, is defined in Eq. 5. For the $3\omega$ method, $u_j$ should be replaced with $\tilde{u}_j$ (Eq. 31).

| | |
|---|---|
| $\mathbf{S_{i,j+1}}(1,1)$ | $\dfrac{\mathbf{S_{i,j}}(1,1)e^{-u_j d_j}}{\mathbf{I_{j,j+1}}(1,1) - \mathbf{S_{i,j}}(1,2)\mathbf{I_{j,j+1}}(2,1)}$ |
| $\mathbf{S_{i,j+1}}(1,2)$ | $\dfrac{\left[\mathbf{S_{i,j}}(1,1)\mathbf{I_{j,j+1}}(2,2)e^{-u_j d_j} - \mathbf{I_{j,j+1}}(1,2)\right]e^{-u_{j+1}d_{j+1}}}{\mathbf{I_{j,j+1}}(1,1) - \mathbf{S_{i,j}}(1,2)\mathbf{I_{j,j+1}}(2,1)}$ |
| $\mathbf{S_{i,j+1}}(2,1)$ | $\mathbf{S_{i,j}}(2,1) + \mathbf{S_{i,j}}(2,2)\mathbf{S_{i,j+1}}(1,1)\mathbf{I_{j,j+1}}(2,1)$ |
| $\mathbf{S_{i,j+1}}(2,2)$ | $\mathbf{S_{i,j}}(2,2)\left[\mathbf{S_{i,j+1}}(1,2)\mathbf{I_{j,j+1}}(2,1) + \mathbf{I_{j,j+1}}(2,2)e^{-u_{j+1}d_{j+1}}\right]$ |

TABLE II. Expressions of normalized coefficients of the temperature fields (Eq. 9), $a_j' = a_j/q_{htr}$ and $b_j' = b_j/q_{htr}$, for the general scenario (Fig. 2). Here

$$\mathbf{H_{j,j+1}} = \begin{bmatrix} \mathbf{M_{j+1}}(1,1) + \mathbf{M_{j+1}}(1,2)\beta e^{-u_{j+1}d_{j+1}} & -\mathbf{M_j}(1,1)\alpha e^{-u_j d_j} - \mathbf{M_j}(1,2) \\ \mathbf{M_{j+1}}(2,1) + \mathbf{M_{j+1}}(2,2)\beta e^{-u_{j+1}d_{j+1}} & -\mathbf{M_j}(2,1)\alpha e^{-u_j d_j} - \mathbf{M_j}(2,2) \end{bmatrix},$$ where

$\mathbf{M_j}$, $u_j$, α and $\beta$ are defined in Eqs. 13, 5, 28 and 29, respectively. If the driving flux is applied on top of the sample (see, for example, the first two schematics in TABLE III), one replaces the subscript $j$ with 0. For the $3\omega$ method, one replaces $u$ with $\tilde{u}$ (Eq. 31), $q_{htr}$ with its Fourier cosine transform, $\tilde{q}_{htr}$, and Eq. 9 with Eq. 32.

| | |
|---|---|
| $a_j' = \dfrac{a_j}{q_{htr}}$ | $\dfrac{\mathbf{M_{j+1}}(1,1) + \mathbf{M_{j+1}}(1,2)\beta e^{-u_{j+1}d_{j+1}}}{\lvert\mathbf{H_{j,j+1}}\rvert}\beta$ |
| $b_j' = \dfrac{b_j}{q_{htr}}$ | $\dfrac{\mathbf{M_{j+1}}(1,1) + \mathbf{M_{j+1}}(1,2)\beta e^{-u_{j+1}d_{j+1}}}{\lvert\mathbf{H_{j,j+1}}\rvert}$ |
| $a_{j+1}' = \dfrac{a_{j+1}}{q_{htr}}$ | $\dfrac{\mathbf{M_j}(1,1)\alpha e^{-u_j d_j} + \mathbf{M_j}(1,2)}{\lvert\mathbf{H_{j,j+1}}\rvert}$ |
| $b_{j+1}' = \dfrac{b_{j+1}}{q_{htr}}$ | $\dfrac{\mathbf{M_j}(1,1)\alpha e^{-u_j d_j} + \mathbf{M_j}(1,2)}{\lvert\mathbf{H_{j,j+1}}\rvert}\beta$ |

TABLE III. Verification of the scattering-matrix framework using well-known solutions, which assume isotropic properties and neglect thermal contact resistance between the film and the substrate, i.e. $R''_{c,film-sub.} = 0$. The general solutions that consider the anisotropic nature of the sample and $R''_{c,film-sub.}$ are given in Section III-C. More details of intermediate results can be found from Table D1 in Appendix D.

| Schematic | **S**-matrix | $\theta_{htr}(x)$ or $\theta_{htr,avg.}$ |
| --- | --- | --- |
| 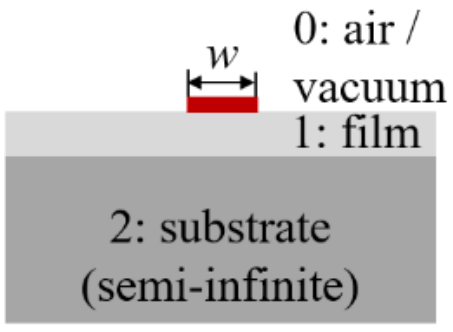  | $\mathbf{S}_{0,0} = \mathbf{S}_{1,1} = \begin{bmatrix} 1 & 0 \\ 0 & 1 \end{bmatrix}$ | $\frac{P_0}{\pi l k_1} K_0\left(\sqrt{i\omega_H/D_1}\,x\right)$<br>(recovers Eq. 1 of Ref. [24]) |
| 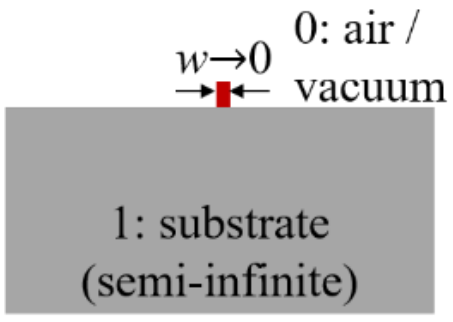  | $\mathbf{S}_{0,0} = \begin{bmatrix} 1 & 0 \\ 0 & 1 \end{bmatrix}$<br>$\mathbf{S}_{1,2} = \begin{bmatrix} \frac{2\tilde{\gamma}_1}{\tilde{\gamma}_1+\tilde{\gamma}_2} e^{-\tilde{u}_1 d_1} & 0 \\ \frac{\tilde{\gamma}_1-\tilde{\gamma}_2}{\tilde{\gamma}_1+\tilde{\gamma}_2} e^{-\tilde{u}_1 d_1} & 0 \end{bmatrix}$ | $\frac{P_0}{\pi l k_2}\left[\frac{1}{2} ln\left(\frac{D_2}{(w/2)^2}\right) + \eta - \frac{1}{2} ln(\omega_H) - i\frac{\pi}{4}\right] + \frac{P_0}{wl}\frac{d_1}{k_1}$<br>(recovers Eqs. 1 and 2 of Ref. [27];<br>see Section III-C-2 for assumptions) |
| 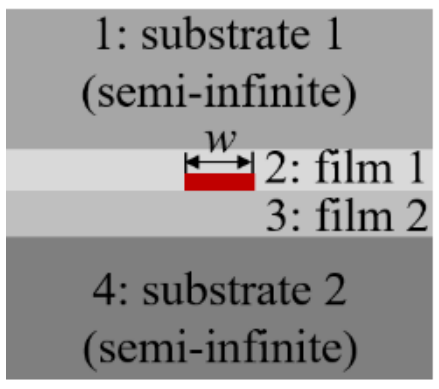  | $\mathbf{S}_{1,2} = \left[0, \frac{\tilde{\gamma}_2-\tilde{\gamma}_1}{\tilde{\gamma}_1+\tilde{\gamma}_2} e^{-\tilde{u}_2 d_2}; 0, \frac{2\tilde{\gamma}_2}{\tilde{\gamma}_1+\tilde{\gamma}_2} e^{-\tilde{u}_2 d_2}\right]$<br>$\mathbf{S}_{3,4} = \left[\frac{2\tilde{\gamma}_3}{\tilde{\gamma}_3+\tilde{\gamma}_4} e^{-\tilde{u}_3 d_3}, 0; \frac{\tilde{\gamma}_3-\tilde{\gamma}_4}{\tilde{\gamma}_3+\tilde{\gamma}_4} e^{-\tilde{u}_3 d_3}, 0\right]$ | $\frac{P_0}{\pi l(k_1+k_4)} \int_0^\infty \frac{1}{\sqrt{\lambda^2 + i\omega_H/D_z}} \frac{sin^2\left(\lambda\frac{w}{2}\right)}{\left(\lambda\frac{w}{2}\right)^2} d\lambda$<br>(recovers Eq. B1 of Ref. [28]<br>assuming 1 and 2 are the same materials,<br>and so are 3 and 4; Boundary Mismatch Assumption) |

Table IV. Input thermal properties (from Ref. [29] if not cited separately) to the FEM models to generate synthetic "experimental data" (symbols in Fig. 4) and those input to the transfer or scattering matrix model to fit for the only free parameter, the cross-plane thermal conductivity of films, $k_{film,z}$.

| Cases in Fig. 4 | Material | Input parameters to FEM model (synthetic "experiments") | | | | Input parameters to transfer or scattering matrix model | | |
|---|---|---|---|---|---|---|---|---|
| | | $C$ [$10^6$ J/m$^3$-K] | $k_x$ [W/m-K] | $k_z$ [W/m-K] | $R''_{c,film-Si}$ [m$^2$-K /W] | $C$ [$10^6$ J/m$^3$-K] | $k_x$ [W/m-K] | $R''_{c,film-Si}$ [m$^2$-K /W] |
| 1 | $SiO_2$ | 1.65 | 1.38 | | $1 \times 10^{-8}$[30] | 1.65 | N/A | 0 |
| | Si | 1.66 | 148 | | | 1.66 | $=k_z$=145$^+$ | |
| 2 | Diamond | 1.78 | 130[43] | 710[43] | $1 \times 10^{-8}$[30] | 1.78 | 143$^*$ | $1.17 \times 10^{-8\#}$ |
| | Si | 1.66 | 148 | | | 1.66 | $=k_z$=148$^+$ | |
| 3 | Diamond | 1.78 | 130[43] | 710[43] | $1 \times 10^{-8}$[30] | 1.78 | 143$^*$ | $1.17 \times 10^{-8\#}$ |
| | Si | 1.66 | 148 | | | 1.66 | $=k_z$=145$^+$ | |

$^+$Obtained by fitting to the synthetic "experimental data" using the slope method[24].

$^*$Assumed to obtain from separate measurement using the microfabricated suspended device method, with representative uncertainty[44].

$^\#$See Appendix G.

TABLE V. Analogy and contrast of key steps to apply the scattering matrix formalism to three popular experimental techniques. Here we assume principal axes of the thermal conductivity tensor are aligned with the natural cartesian coordinate system defined by the heater line and sample surface. The key is to obtain the complex wavevector, $u$ (or, $\tilde{u}$ for multi-dimensional cases). Here $D_z = k_z/C$, $n_{xz} = k_x/k_z$, and $n_{rz} = k_r/k_z$

| | 1D Planar Heating | $3\omega$ Method[6-7] | TDTR[10-13]/FDTR[14-15] |
|---|---|---|---|
| Governing equation (GE) | $C\frac{\partial T}{\partial t} = k_z\frac{\partial^2 T}{\partial z^2}$ | $C\frac{\partial T}{\partial t} = k_x\frac{\partial^2 T}{\partial x^2} + k_z\frac{\partial^2 T}{\partial z^2}$ | $C\frac{\partial T}{\partial t} = k_r\left(\frac{\partial^2 T}{\partial r^2} + \frac{1}{r}\frac{\partial T}{\partial r}\right) + k_z\frac{\partial^2 T}{\partial z^2}$ |
| Time-independent GE | $\frac{i\omega_H}{D_z}\theta = k_z\frac{\partial^2 \theta}{\partial z^2}$ | $\frac{i\omega_H}{D_z}\theta = n_{xz}\frac{\partial^2 \theta}{\partial x^2} + \frac{\partial^2 \theta}{\partial z^2}$ | $\frac{i\omega_H}{D_z}\theta = n_{rz}\left(\frac{\partial^2 \theta}{\partial r^2} + \frac{1}{r}\frac{\partial \theta}{\partial r}\right) + \frac{\partial^2 \theta}{\partial z^2}$ |
| Integral transform | | Fourier cosine transform<br>$\int_0^\infty \theta(x,z)\cos(\lambda x)\,dx$ | Hankel transform<br>$\int_0^\infty \theta(r,z)J_0(2\pi\lambda r)r\,dr$ |
| GE in transformed space | | $\frac{i\omega_H}{D_z}\tilde{\theta} = -n_{xz}\lambda^2\tilde{\theta} + \frac{\partial^2\tilde{\theta}}{\partial z^2}$ | $\frac{i\omega_H}{D_z}\tilde{\theta} = -n_{rz}(2\pi\lambda)^2\tilde{\theta} + \frac{\partial^2\tilde{\theta}}{\partial z^2}$ |
| Complex wave vector | $u = \sqrt{\frac{i\omega_H}{D_z}}$ | $\tilde{u} = \sqrt{n_{xz}\lambda^2 + \frac{i\omega_H}{D_z}}$ | $\tilde{u} = \sqrt{n_{rz}(2\pi\lambda)^2 + \frac{i\omega_H}{D_z}}$ |

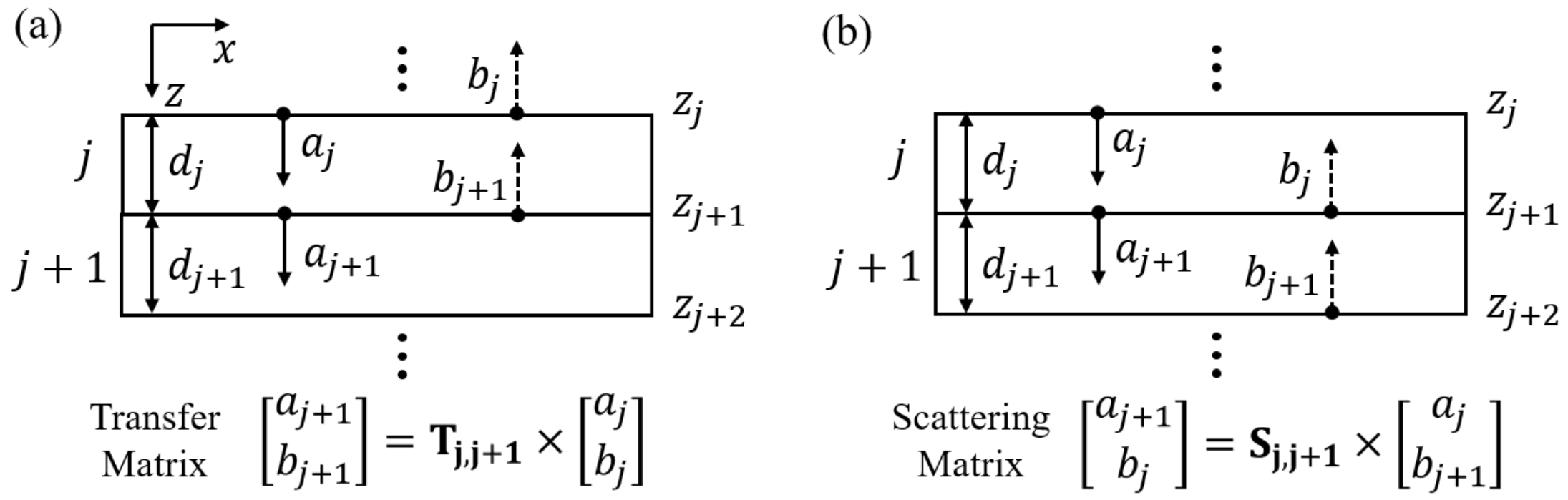

Fig. 1. Comparison between (a) transfer matrix and (b) scattering matrix formalism. The former uses two coefficients, $a_j$ and $b_j$, at the same reference point, $z = z_j$, to represent the temperature field (see Eq. 4 of the main text) and the corresponding heat flux in layer *j*. The temperature field and flux in layer *j*+1 are linked to those in layer *j* using the transfer matrix. Although straightforward, this transfer matrix formalism is numerically unstable, as discussed in session II-A of the main text. To circumvent this problem, the scattering matrix formalism in this work changes the reference point from $z = z_j$ to $z = z_{j+1}$, and skews the input and output of the transfer matrix formalism.

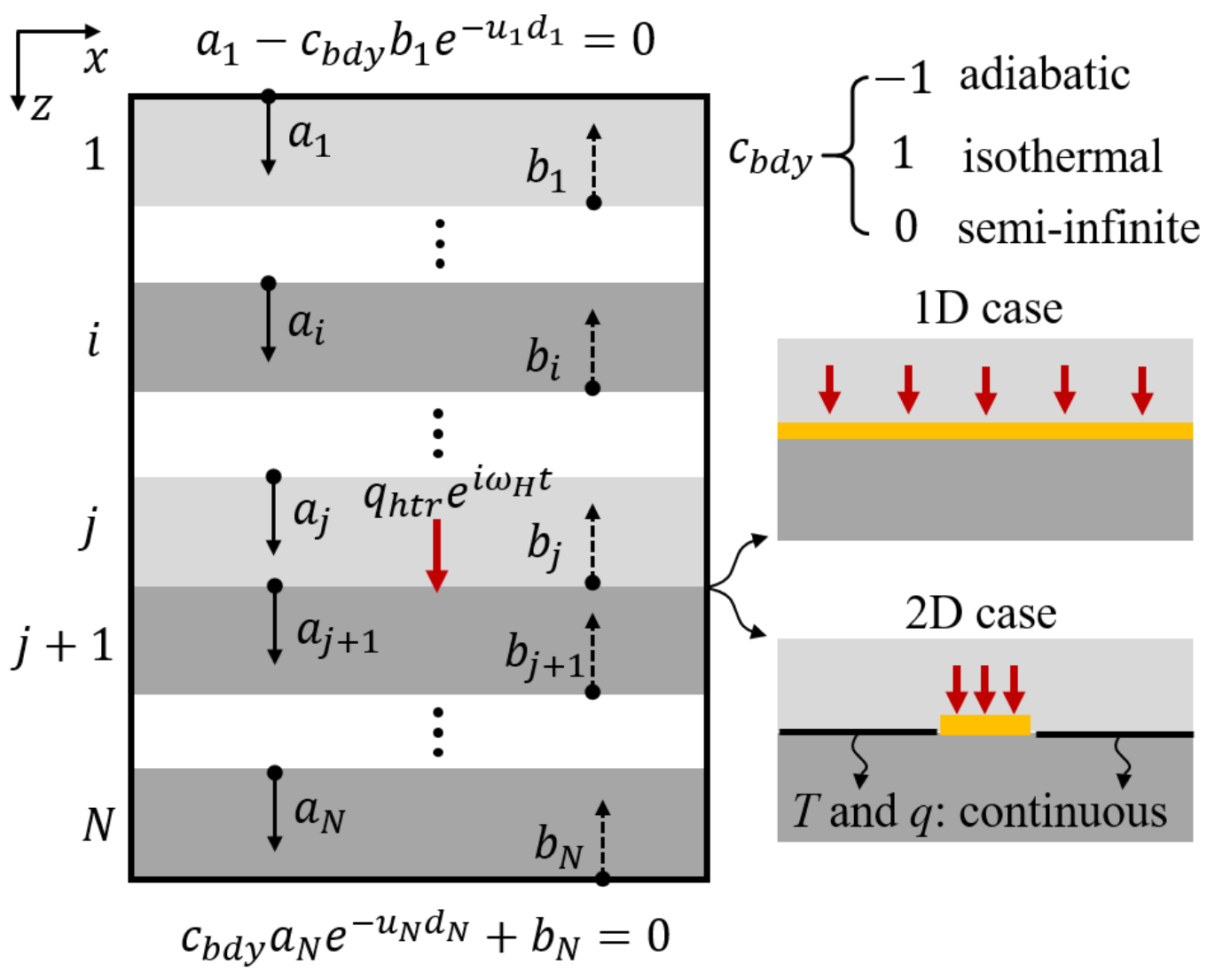

Fig. 2. Schematic of the general framework: boundary conditions and the driving flux ($q_{htr} e^{i\omega_H t}$) exerted at the interface between layers $j$ and $j+1$. The thermal contact resistance between adjacent layers can also be included in the framework (see Eqs. 17-18).

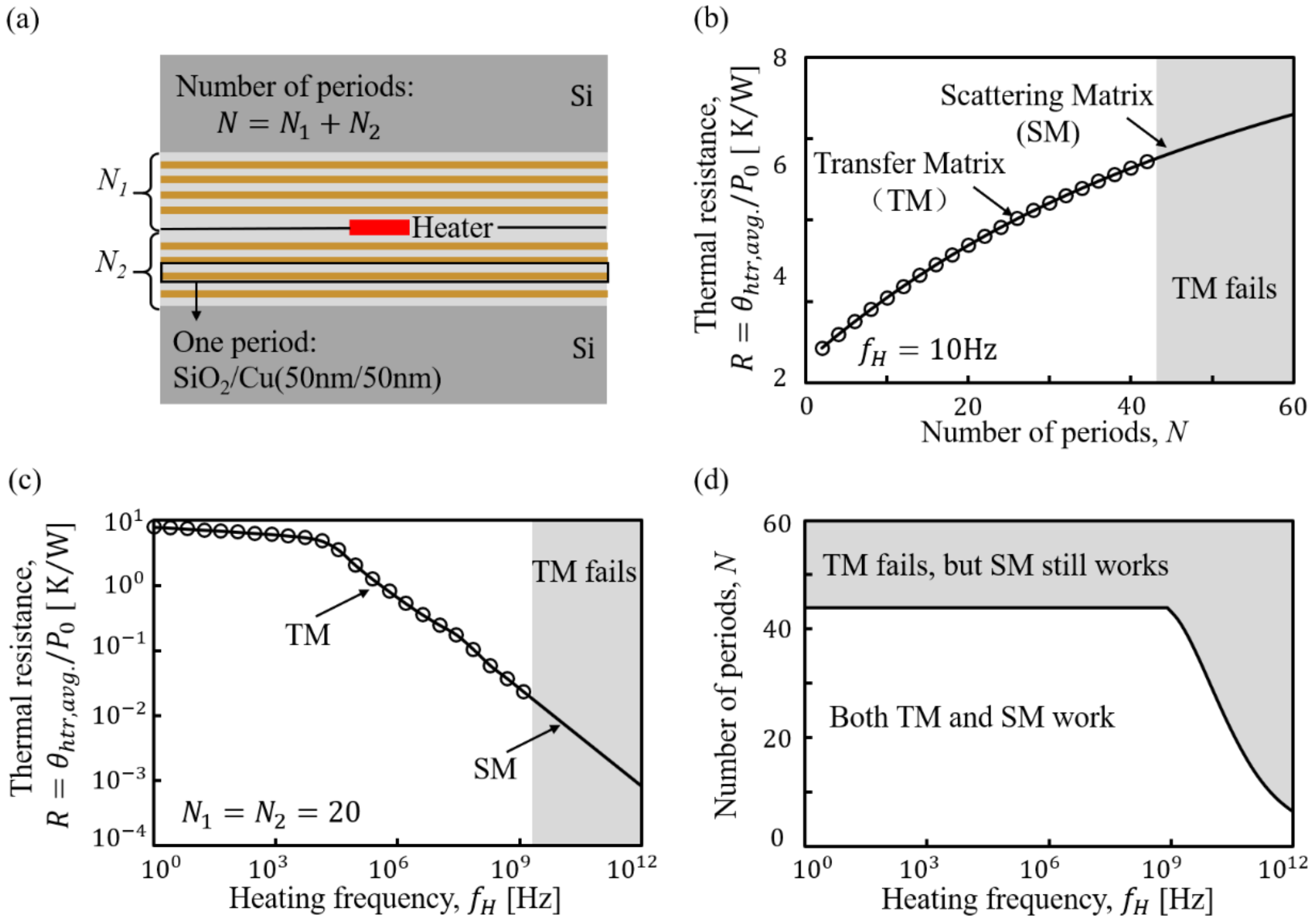

Fig. 3. Demonstration of the numerical stability of the scattering matrix formalism. (a) Schematic: Periodic heating of alternating $SiO_2$-Cu structures with a Si substrate on both sides. (b) $R$ $(= \theta_{htr,avg.}/P_0)$ vs. $N$ relation at a fixed heating frequency, $f_H$ = 10 Hz. (c) $R$ vs. $f_H$ relation at a fixed total number of periods, $N$ = 40. (d) Regime map: While the transfer matrix framework fails at a critical $N_c$ or $f_{H,c}$, the scattering matrix methodology is robust everywhere in the $f_H$-$N$ plane.

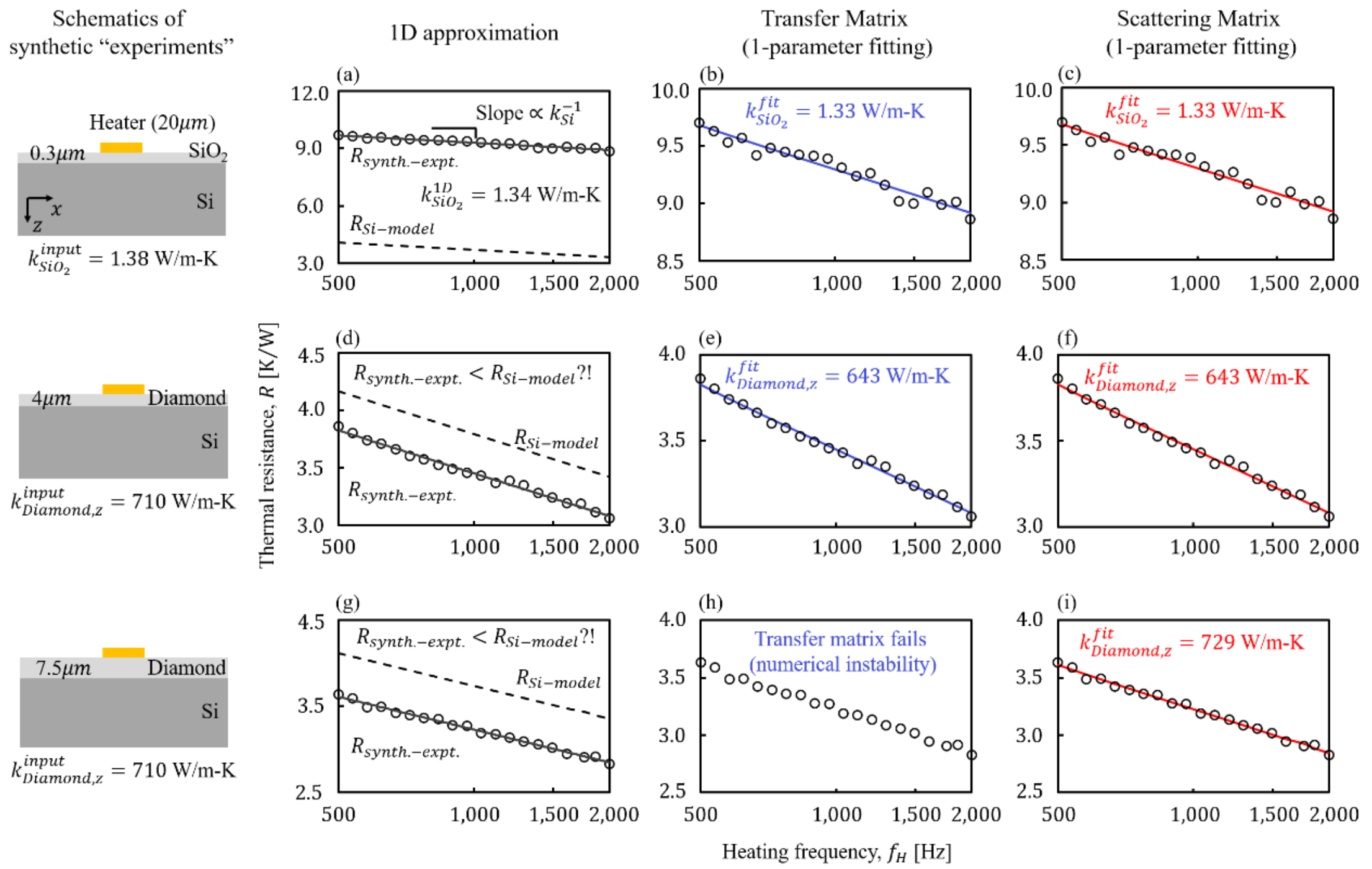

Fig. 4. Synthetic "experiments" to demonstrate the merit of the scattering matrix formalism under certain circumstances. Representative "Experimental" data points with $\pm 1\%$ of "experimental noise" (open circles) are generated using a 3D FEM model (see Table IV for a summary of input thermal properties). To extract the cross-plane thermal conductivity of the film, $k_{film,z}$, three methods (columns) are applied to three scenarios (rows). $k_{Si}$ is obtained using the slope method[24] for all three sets of "experiments." While the 1D approximation ($1^{st}$ column) method and the transfer matrix methodology ($2^{nd}$ column) successively fail (Figs. 4d, 4g and 4h), the scattering matrix formalism (last column) remains reliable. See Appendix G for more details on the sensitivity and uncertainty analyses.

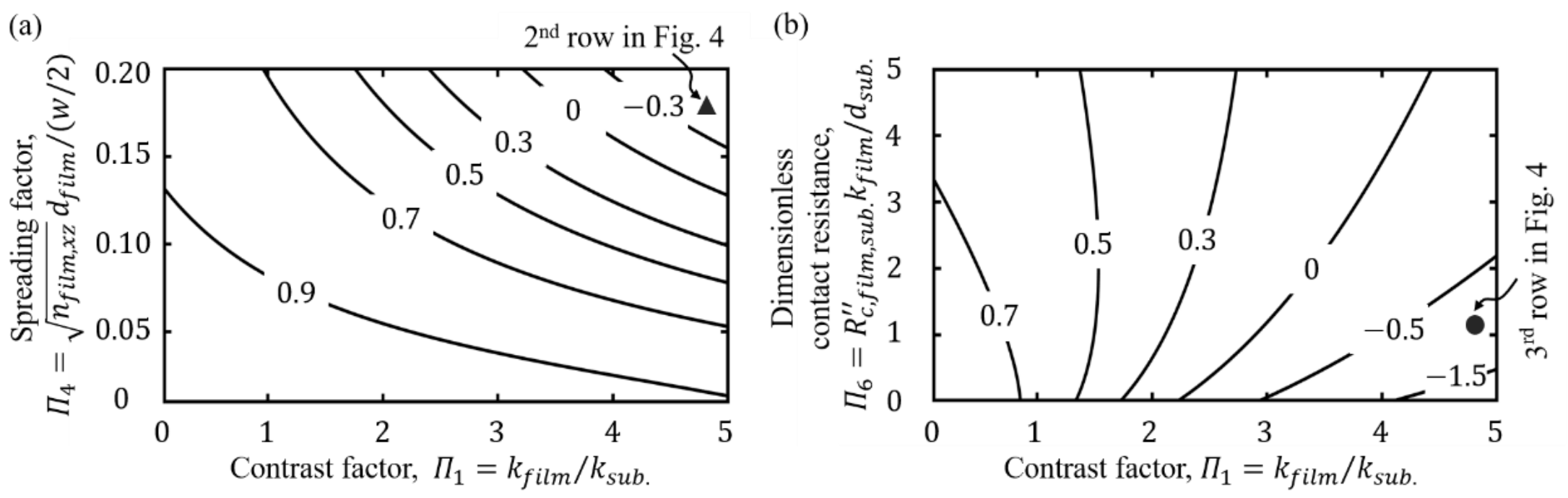

Fig. 5. Quantitative interpretation to the negative $R_{film}$ in Figs. 4d and 4g caused by the 1D approximation (Eq. 50). Contour plots show the dimensionless scaling factor (Eq. E1 in Appendix E), $s = \theta_{film}/\theta_{film,1D} = R_{film}/R_{1D}$, as a function of (a) the contrast factor, $\Pi_1 = k_{film,z}/k_{sub.,z}$ , and the spreading factor, $\Pi_4 = \sqrt{k_{film,x}/k_{film,z}}\frac{d_{film}}{w/2}$ ; (b) $\Pi_1$ and the dimensionless contact resistance, $\Pi_6 = R^{''}_{c,film-sub.} k_{film}/d_{film}$. $s = 1$ means that the 1D conduction formula (Eq. 50) is exact. Both scenarios 2 and 3 in Fig. 4 deviates significantly from the 1D approximation mainly due to the large $\Pi_1$.

APPENDICES

**A Scattering Matrix Formalism to Model Periodic Heat Diffusion in Stratified Solid Media**

**Tao Li and Zhen Chen***

Jiangsu Key Laboratory for Design & Manufacture or Micro/Nano Biomedical Instruments, School of Mechanical Engineering, Southeast University, Nanjing 210096, China

* correspondence should be addressed to: zhenchen@seu.edu.cn

APPENDIX A: **Nomenclature**

| | |
|---|---|
| **Latin letters** | |
| $a$, $b$ | Coefficients of the temperature field, see Eq. 4 and Fig. 1 |
| $c$ | A factor distinguishing different boundary conditions, see Eq. 23 |
| $d$ | Layer thickness |
| $e$ | Thermal effusivity |
| $f$ | Frequency of the driving electrical current |
| $f_H$ | Frequency of the Joule heating |
| $i$ | Unit imaginary number, $\sqrt{-1}$ |
| $j$ | Layer index |
| $k$ | Thermal conductivity |
| $l$ | Heater length |
| $n$ | Anisotropy ratio of the thermal conductivity |
| $P_0$ | Root-mean-square value of the Joule heating power, [W] |
| $q$ | Heat flux in real space, [Wm$^{-2}$] |
| $\tilde{q}$ | Fourier cosine transform of $q$ |
| $r$ | Radial component of the cylindrical coordinate |
| $s$ | Dimensionless scaling factor, see Eq. 49 |
| $t$ | Time |
| $u$ | Magnitude of the complex thermal wave vector (1D), see Eq. 5 |
| $\tilde{u}$ | Magnitude of the complex thermal wave vector (2D), see Eq. 31 |
| $w$ | Heater width |
| $x$, $z$ | Components of the Cartesian coordinate, see Fig. 1 |
| $B$ | Dimensionless group containing $R_c^{''}$, $\tilde{u}$, and $\tilde{\gamma}$, see Eq. 47 |
| $C$ | Volumetric heat capacity |
| $D$ | Thermal diffusivity |
| $K_0$ | Modified Bessel functions of the second kind |
| $N$ | Total layer number |
| $R$ | Thermal resistance |
| $T$ | Spatial and temporal temperature field |
| $\mathbf{I}$ | Interface matrix, see Eq. 16 |
| $\mathbf{I}'$ | Interface matrix involving thermal contact resistance, see Eq. 17 |
| $\mathbf{M}$ | Material matrix, see Eq. 13 |
| $\mathbf{M}'$ | Material matrix involving thermal contact resistance, see Eq. 18 |
| $\mathbf{E}$ | Thermal effusivity matrix |
| $\mathbf{S}$ | Scattering matrix |
| $\mathbf{T}$ | Transfer matrix |
| | |
| **Greek letters** | |
| $\alpha$ | Coefficient bridging $a_j$ and $b_j$, see Eq. 26 |
| $\beta$ | Coefficient bridging $a_{j+1}$ and $b_{j+1}$, see Eq. 27 |
| $\gamma$ | Product of $k_j$ and $u_j$ |
| $\tilde{\gamma}$ | Product of $k_j$ and $\tilde{u}_j$ |
| $\delta$ | Dirac delta function |
| $\epsilon$ | Anisotropy ratio of the thermal conductivity |
| $\theta$ | Spatial temperature field |
| $\tilde{\theta}$ | Fourier cosine transform of $\theta$ |
| $\omega$ | Angular frequency |
| $\Pi_1$-$\Pi_6$ | Dimensionless groups, see Eqs. 49 and E1 |
| $\mathbf{\Phi}$ | Phase matrix, see Eq. 8 |

**Subscripts and Superscripts**

| | |
|---|---|
| $''$ | Per unit area |
| $avg.$ | Average |
| $bdy$ | Boundary |
| c | Contact |
| $f$ | Film |
| $htr$ | Heater |
| $H$ | Heating |
| $i$ | Layer index |
| $j$ | Layer index |
| $N$ | Layer index |
| $ref$ | Reference point |
| $sub.$ | Substrate |
| $1,j$ | From layer 1 to layer $j$ |
| $j,z$ | Along $z$-direction inside layer $j$ |
| $j,xz$ | Thermal properties along $x$-direction normalized by those along $z$-direction inside layer $j$ |
| $xz$ | Thermal properties along $x$-direction normalized by those along $z$-direction |

APPENDIX B: Derivation of the recursive relations in Table I of the main text

Once $\mathbf{S_{i,j}}$ is known, one can recursively obtain $\mathbf{S_{i,j+1}}$ using Eqs. 11 and 15. First, from Eq. 11 one has

$$a_j = \mathbf{S_{i,j}}(1,1)a_i + \mathbf{S_{i,j}}(1,2)b_j, \tag{B1}$$

$$b_i = \mathbf{S_{i,j}}(2,1)a_i + \mathbf{S_{i,j}}(2,2)b_j. \tag{B2}$$

Replacing the subscript *j* with *j*+1 in Eqs. B1 and B2, one obtains

$$a_{j+1} = \mathbf{S_{i,j+1}}(1,1)a_i + \mathbf{S_{i,j+1}}(1,2)b_{j+1}, \tag{B3}$$

$$b_i = \mathbf{S_{i,j+1}}(2,1)a_i + \mathbf{S_{i,j+1}}(2,2)b_{j+1}. \tag{B4}$$

Second, from Eq. 15 one has

$$a_j e^{-u_j d_j} = \mathbf{I_{j,j+1}}(1,1)a_{j+1} + \mathbf{I_{j,j+1}}(1,2)b_{j+1}e^{-u_{j+1}d_{j+1}}, \tag{B5}$$

$$b_j = \mathbf{I_{j,j+1}}(2,1)a_{j+1} + \mathbf{I_{j,j+1}}(2,2)b_{j+1}e^{-u_{j+1}d_{j+1}}. \tag{B6}$$

Last, substituting Eqs. B5 and B6 into Eq. B1 to eliminate $a_j$ and $b_j$, one obtains

$$a_{j+1} = \frac{\mathrm{S_{i,j}}(1,1)e^{-u_j d_j}}{\mathbf{I_{j,j+1}}(1,1)-\mathbf{S_{i,j}}(1,2)\mathbf{I_{j,j+1}}(2,1)}a_i + \frac{\left[\mathbf{S_{i,j}}(1,1)\mathbf{I_{j,j+1}}(2,2)e^{-u_j d_j}-\mathbf{I_{j,j+1}}(1,2)\right]e^{-u_{j+1}d_{j+1}}}{\mathbf{I_{j,j+1}}(1,1)-\mathbf{S_{i,j}}(1,2)\mathbf{I_{j,j+1}}(2,1)}b_{j+1}. \tag{B7}$$

Comparing Eqs. B3 and B7, one obtains

$$\mathbf{S_{i,j+1}}(1,1) = \frac{\mathbf{S_{i,j}}(1,1)e^{-u_j d_j}}{\mathbf{I_{j,j+1}}(1,1)-\mathbf{S_{i,j}}(1,2)\mathbf{I_{j,j+1}}(2,1)}, \tag{B8}$$

$$\mathbf{S_{i,j+1}}(1,2) = \frac{\left[\mathbf{S_{i,j}}(1,1)\mathbf{I_{j,j+1}}(2,2)e^{-u_j d_j}-\mathbf{I_{j,j+1}}(1,2)\right]e^{-u_{j+1}d_{j+1}}}{\mathbf{I_{j,j+1}}(1,1)-\mathbf{S_{i,j}}(1,2)\mathbf{I_{j,j+1}}(2,1)}. \tag{B9}$$

Likewise, substituting Eqs. B5 and B6 into Eq. B2 to eliminate $a_{j+1}$ and $b_j$, one obtains

$$b_i = \left[\mathbf{S_{i,j}}(2,1) + \mathbf{S_{i,j}}(2,2)\mathbf{S_{i,j+1}}(1,1)\mathbf{I_{j,j+1}}(2,1)\right]a_i,$$

$$+\mathbf{S_{i,j}}(2,2)\left[\mathbf{S_{i,j+1}}(1,2)\mathbf{I_{j,j+1}}(2,1) + \mathrm{I_{j,j+1}}(2,2)e^{-u_{j+1}d_{j+1}}\right]b_{j+1}. \tag{B10}$$

Comparing Eqs. B4 and B210, one obtains

$$\mathbf{S_{i,j+1}}(2,1) = \mathbf{S_{i,j}}(2,1) + \mathbf{S_{i,j}}(2,2)\mathbf{S_{i,j+1}}(1,1)\mathbf{I_{j,j+1}}(2,1), \tag{B11}$$

$$\mathbf{S_{i,j+1}}(2,2) = \mathbf{S_{i,j}}(2,2)\left[\mathbf{S_{i,j+1}}(1,2)\mathbf{I_{j,j+1}}(2,1) + \mathbf{I_{j,j+1}}(2,2)e^{-u_{j+1}d_{j+1}}\right]. \tag{B12}$$

Therefore, Eqs. B8, B9, B11, and B12 give the recursive relations to obtain $\mathbf{S_{i,j+1}}$ from $\mathbf{S_{i,j}}$, which are summarized in Table I.

APPENDIX C: The general case of an anisotropic thermal conductivity tensor with finite off-diagonal elements

In session III-A of the main text, we assume the thermal conductivity tensor is diagonalized for simplicity. In this appendix, we briefly outline the key steps for the general case of a non-diagonalized anisotropic thermal conductivity tensor. We start from the general form of the 2D diffusion equation,

$$C\frac{\partial T}{\partial t} = k_{xx}\frac{\partial^2 T}{\partial x^2} + 2k_{xz}\frac{\partial^2 T}{\partial x \partial z} + k_{zz}\frac{\partial^2 T}{\partial z^2}. \tag{C1}$$

Substituting Eq. 2 into Eq. C1, one obtains

$$\frac{i\omega_H}{D_{zz}}\theta = \epsilon_{xx}\frac{\partial^2 \theta}{\partial x^2} + 2\epsilon_{xz}\frac{\partial^2 \theta}{\partial x \partial z} + \frac{\partial^2 \theta}{\partial z^2}, \tag{C2}$$

where $\epsilon_{xx} = k_{xx}/k_{zz}$, $\epsilon_{xz} = k_{xz}/k_{zz}$ and $D_{zz} = C/k_{zz}$, Performing the Fourier cosine transform (along the *x*-axis) to Eq. C2, one obtains

$$\frac{i\omega_H}{D_{zz}}\tilde{\theta} = -\lambda^2\epsilon_{xx}\tilde{\theta} + 2i\lambda\epsilon_{xz}\frac{\partial\tilde{\theta}}{\partial z} + \frac{\partial^2\tilde{\theta}}{\partial z^2}. \tag{C3}$$

The general solution of Eq. C3 is

$$\tilde{\theta}_j(z) = a_j\big|_{z=z_j} e^{-\tilde{u}_j(z-z_j)} + b_j\big|_{z=z_j+d_j} e^{-\tilde{u}_j(z_j+d_j-z)}, \tag{C4}$$

As compared to the simpler case where the thermal conductivity tensor is diagonalized (Eq. 31 of the main text), the complex wave vector is updated to

$$\tilde{u}_j = i\lambda\epsilon_{xz} + \sqrt{(\epsilon_{xx} - \epsilon_{xz}^2)\lambda^2 + \frac{i\omega_H}{D_{zz}}}. \tag{C5}$$

According to Fourier's law, the heat flux in the *z* direction is

$$q_j(z) = -k_{zz}\frac{\partial\theta}{\partial z} - k_{xz}\frac{\partial\theta}{\partial x}, \tag{C6}$$

where $k_{xz} = k_{zx}$ by reciprocity relation[1]. Taking the Fourier cosine transform (along the *x*-axis) to Eq. C6, one obtains

$$\tilde{q}_j(z) = -k_{zz}\frac{\partial\tilde{\theta}}{\partial z} - i\lambda k_{xz}\tilde{\theta}. \tag{C7}$$

Therefore, Eq. 13 of the main text is updated to

$$\mathbf{M_j} = \begin{bmatrix} 1 & 1 \\ \tilde{\gamma}_{j,zz} - i\lambda k_{j,xz} & -\tilde{\gamma}_{j,zz} - i\lambda k_{j,xz} \end{bmatrix}, \tag{C8}$$

where

$$\tilde{\gamma}_{j,zz} = k_{j,zz}\tilde{u}_j \tag{C9}$$

In summary, to take into account the most general scenario where the anisotropic thermal conductivity tensor has finite off-diagonal elements, one can simply update $\tilde{u}_j$, $\mathbf{M_j}$, and $\tilde{\gamma}_j$ from Eqs. 31, 13, and 14 of the main text to Eqs. C5, C8, and C9, respectively.

APPENDIX D: More details of implementing the scattering matrix method

In Table III and Section III-C of the main text, we outline the S-matrix and the temperature of the heater of well-known scenarios. In this appendix, we show more details of the intermediate results, including the M-matrix (Eq. 13), the I-matrix (Eq. 16), the S-matrix (Table I), the coefficients $\alpha$ (Eq. 28) and $\beta$ (Eq. 29), the normalized coefficients of the temperature of the heater $a'_{j+1}$ and $b'_{j+1}$ (Table II), and the general expression of the temperature of the heater (Eqs. 32 and 33) that takes into account the anisotropic nature of the sample. Note here we assume perfect thermal contact between adjacent layers, i.e. $R_c'' = 0$.

Table D1. More intermediate results to supplement Table III of the main text. Here $\tilde{u}_j = \sqrt{n_{j,xz}\lambda^2 + \frac{i\omega_H}{D_{j,z}}}$ (Eq. 31), $\tilde{\gamma}_j = k_j\tilde{u}_j$ (Eq. 14), and $B_{ij}^{\pm} = \frac{1}{2\tilde{\gamma}_i}(\tilde{\gamma}_i \pm \tilde{\gamma}_j)e^{\pm\tilde{u}_i d_i}$ (Eq. 47).

| | 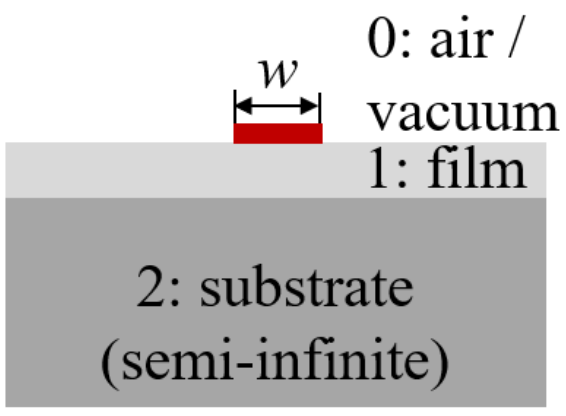 | 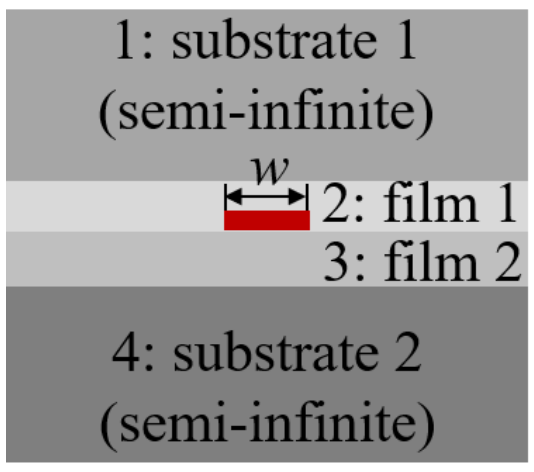 |
|---|---|---|
| M-matrix | $\mathbf{M_0} = [1,1;0,0], \mathbf{M_1} = [1,1;\tilde{\gamma}_1,-\tilde{\gamma}_1]$ <br> $\mathbf{M_2} = [1,1;\tilde{\gamma}_2,-\tilde{\gamma}_2]$ | $\mathbf{M_1} = [1,1;\tilde{\gamma}_1,-\tilde{\gamma}_1], \mathbf{M_2} = [1,1;\tilde{\gamma}_2,-\tilde{\gamma}_2]$ <br> $\mathbf{M_3} = [1,1;\tilde{\gamma}_3,-\tilde{\gamma}_3], \mathbf{M_4} = [1,1;\tilde{\gamma}_4,-\tilde{\gamma}_4]$ |
| I-matrix | $\mathbf{I_{1,2}} = \frac{1}{2}\left[1+\frac{\tilde{\gamma}_2}{\tilde{\gamma}_1}, 1-\frac{\tilde{\gamma}_2}{\tilde{\gamma}_1}; 1-\frac{\tilde{\gamma}_2}{\tilde{\gamma}_1}, 1+\frac{\tilde{\gamma}_2}{\tilde{\gamma}_1}\right]$ | $\mathbf{I_{1,2}} = \frac{1}{2}\left[1+\frac{\tilde{\gamma}_2}{\tilde{\gamma}_1}, 1-\frac{\tilde{\gamma}_2}{\tilde{\gamma}_1}; 1-\frac{\tilde{\gamma}_2}{\tilde{\gamma}_1}, 1+\frac{\tilde{\gamma}_2}{\tilde{\gamma}_1}\right]$ <br> $\mathbf{I_{3,4}} = \frac{1}{2}\left[1+\frac{\tilde{\gamma}_4}{\tilde{\gamma}_3}, 1-\frac{\tilde{\gamma}_4}{\tilde{\gamma}_3}; 1-\frac{\tilde{\gamma}_4}{\tilde{\gamma}_3}, 1+\frac{\tilde{\gamma}_4}{\tilde{\gamma}_3}\right]$ |
| S-matrix | $\mathbf{S_{0,0}} = [1,0;0,1]$ <br> $\mathbf{S_{1,2}} = \left[\frac{2\tilde{\gamma}_1}{\tilde{\gamma}_1+\tilde{\gamma}_2}e^{-\tilde{u}_1 d_1}, 0; \frac{\tilde{\gamma}_1-\tilde{\gamma}_2}{\tilde{\gamma}_1+\tilde{\gamma}_2}e^{-\tilde{u}_1 d_1}, 0\right]$ | $\mathbf{S_{1,2}} = \left[0, \frac{\tilde{\gamma}_2-\tilde{\gamma}_1}{\tilde{\gamma}_1+\tilde{\gamma}_2}e^{-\tilde{u}_2 d_2}; 0, \frac{2\tilde{\gamma}_2}{\tilde{\gamma}_1+\tilde{\gamma}_2}e^{-\tilde{u}_2 d_2}\right]$ <br> $\mathbf{S_{3,4}} = \left[\frac{2\tilde{\gamma}_3}{\tilde{\gamma}_3+\tilde{\gamma}_4}e^{-\tilde{u}_3 d_3}, 0; \frac{\tilde{\gamma}_3-\tilde{\gamma}_4}{\tilde{\gamma}_3+\tilde{\gamma}_4}e^{-\tilde{u}_3 d_3}, 0\right]$ |
| $\alpha$ | $\mathbf{S_{0,0}}(1,2) = 0$ | $\mathbf{S_{1,2}}(1,2) = \frac{\tilde{\gamma}_2-\tilde{\gamma}_1}{\tilde{\gamma}_1+\tilde{\gamma}_2}e^{-\tilde{u}_2 d_2}$ |
| $\beta$ | $\mathbf{S_{1,2}}(2,1) = \frac{\tilde{\gamma}_1-\tilde{\gamma}_2}{\tilde{\gamma}_1+\tilde{\gamma}_2}e^{-\tilde{u}_1 d_1}$ | $\mathbf{S_{3,4}}(2,1) = \frac{\tilde{\gamma}_3-\tilde{\gamma}_4}{\tilde{\gamma}_3+\tilde{\gamma}_4}e^{-\tilde{u}_3 d_3}$ |
| $\tilde{q}_h$ | $\frac{P_0}{2l}\frac{\sin\left(\lambda\frac{w}{2}\right)}{\left(\lambda\frac{w}{2}\right)}$ | $\frac{P_0}{2l}\frac{\sin\left(\lambda\frac{w}{2}\right)}{\left(\lambda\frac{w}{2}\right)}$ |
| $a'_{j+1}$ | $\frac{1}{\tilde{\gamma}_1}\frac{(\tilde{\gamma}_1+\tilde{\gamma}_2)}{(\tilde{\gamma}_1+\tilde{\gamma}_2)-(\tilde{\gamma}_1-\tilde{\gamma}_2)e^{-2\tilde{u}_1 d_1}}$ | $\frac{(B_{21}^- + B_{21}^+)B_{34}^+}{\tilde{\gamma}_3(B_{34}^+ - B_{34}^-)(B_{21}^- + B_{21}^+) - \tilde{\gamma}_2(B_{34}^+ + B_{34}^-)(B_{21}^- - B_{21}^+)}$ |
| $b'_{j+1}$ | $\frac{1}{\tilde{\gamma}_1}\frac{(\tilde{\gamma}_1-\tilde{\gamma}_2)e^{-2\tilde{u}_1 d_1}}{(\tilde{\gamma}_1+\tilde{\gamma}_2)-(\tilde{\gamma}_1-\tilde{\gamma}_2)e^{-2\tilde{u}_1 d_1}}$ | $\frac{(B_{21}^{-}+B_{21}^{+})B_{34}^{-}}{\tilde{\gamma}_3(B_{34}^{+}-B_{34}^{-})(B_{21}^{-}+B_{21}^{+})-\tilde{\gamma}_2(B_{34}^{+}+B_{34}^{-})(B_{21}^{-}-B_{21}^{+})}$ |
| $\theta_{htr,avg}$ | $\frac{P_0}{\pi l}\int_0^\infty \frac{1}{\gamma_1}\frac{B_{12}^{+}+B_{12}^{-}}{B_{12}^{+}-B_{12}^{-}}\frac{\sin^2\left(\lambda\frac{w}{2}\right)}{\left(\lambda\frac{w}{2}\right)^2}d\lambda$ | $\frac{P_0}{\pi l}\int_0^\infty \frac{1}{\tilde{\gamma}_2\frac{(B_{21}^{+}-B_{21}^{-})}{(B_{21}^{-}+B_{21}^{+})}+\tilde{\gamma}_3\frac{(B_{34}^{+}-B_{34}^{-})}{(B_{34}^{+}+B_{34}^{-})}}\frac{\sin^2\left(\lambda\frac{w}{2}\right)}{\left(\lambda\frac{w}{2}\right)^2}d\lambda$ |

APPENDIX E: Dimensionless scaling factor that evaluates the 1D approximation

The explicit expression of the dimensionless scaling factor of Eq. 49 is

$$s=\frac{2}{\pi(1+\Pi_6)}\int_0^\infty [1-\left(\Pi_1\frac{\sqrt{i\Pi_2^2+\Pi_4^2\lambda^2}}{\sqrt{i\Pi_3^2+\Pi_5^2\lambda^2}}\right)^2\frac{(1+\frac{\Pi_6}{\Pi_1}\sqrt{i\Pi_3^2+\Pi_5^2\lambda^2})tanh\left(\sqrt{i\Pi_2^2+\Pi_4^2\lambda^2}\right)}{tanh\left(\sqrt{i\Pi_2^2+\Pi_4^2\lambda^2}\right)+\Pi_6\sqrt{i\Pi_2^2+\Pi_4^2\lambda^2}}]\times$$

$$\frac{tanh\left(\sqrt{i\Pi_2^2+\Pi_4^2\lambda^2}\right)+\Pi_6\sqrt{i\Pi_2^2+\Pi_4^2\lambda^2}}{\left[1+\Pi_1\frac{\sqrt{i\Pi_2^2+\Pi_4^2\lambda^2}}{\sqrt{i\Pi_3^2+\Pi_5^2\lambda^2}}tanh\left(\sqrt{i\Pi_2^2+\Pi_4^2\lambda^2}\right)+\Pi_6\sqrt{i\Pi_2^2+\Pi_4^2\lambda^2}tanh\left(\sqrt{i\Pi_2^2+\Pi_4^2\lambda^2}\right)\right]\sqrt{i\Pi_2^2+\Pi_4^2\lambda^2}}\frac{sin^2(\lambda)}{\lambda^2}d\lambda, \quad \text{(E1)}$$

where the dimensionless groups are defined as $\Pi_1=k_{film,z}/k_{sub.,z}$ , $\Pi_2=d_{film}/L_{penetr.,film}$ , $\Pi_3=d_{film}/L_{penetr.,sub.}$ , $\Pi_4=\sqrt{n_{film,xz}}\frac{d_{film}}{w/2}$ , $\Pi_5=\sqrt{n_{sub,xz}}\frac{d_{film}}{w/2}$, and $\Pi_6=R''_{c,film-sub.}k_{film}/d_{film}$. Here the penetration depth is computed using Eq. 44. Under the approximation of $\max(\Pi_1,\Pi_2,\Pi_4,\Pi_6)\ll 1$ and $\min(\Pi_3,\Pi_5)\gg 1$, and assuming isotropic properties of the sample, one reduces Eq. 48 to Eq. 51, the well-known series-resistor solution.

APPENDIX F: Quantifying the error of Eq. 45

The widely used heat transfer model (Eq. 45) of the $3\omega$ method assumes semi-infinite substrate heated by a narrow heater. This assumption requires
(1) $L_{penetr.} \ll d_{sub.}$ and
(2) $L_{penetr.} \gg w/2$,

where $L_{penetr.} = \sqrt{D_{sub.}/2\pi f_H}$ (Eq. 44). These requirements set the proper frequency range of the measurements.

An open question here is how big of the substrate is big enough? Likewise, how narrow of the heater line is narrow enough? A common practice is to give an empirical safe factor, e.g. $L_{penetr.} \leq \frac{1}{5}d_{sub.}$ and $L_{penetr.} \geq 5 \times (w/2)$ [45]. Dames bridges these safe factors to their corresponding error tolerance[46].

To answer the open question above systematically, we justify the applicable regime of Eq. 45 using the general solution, Eq. 35. The latter is not constrained by the two requirements above. Here we define a relative error, $err. = (R - R_{semi-inf.})/R_{semi-inf.}$, where $R_{semi-inf.}$ is the thermal resistance defined by Eq. 44, and $R$ is the thermal resistance (Eq. 35) corresponding to the isothermal (solid line) and the adiabatic (dashed line) boundary conditions, respectively. For example, Fig. F1a reads that $d_{Sub.}/L_{penetr.} > 2.40$ ensures $err. < \pm 5\%$. Likewise, Fig. F1b suggests that $L_{penetr.}/(w/2) > 1.35$ ensures $err. < \pm 5\%$. We note here in Fig. E1b the far-side boundary condtion does not matter any more, and thus we show only the result of the isothermal boundary condition.

In practical measurements in a cryostat where a specific temperature point is set and mornitored using a thermocouple anchored near the chip carrier mounted on the copper "cold finger", one might think an isothermal boundary condition for the bottom surface of the sample describes the reality most properly. However, this isothermal boundary condition requires an infinite large thermal conductivity of the "cold finger" and the ceramic chip carrier. This requirement cannot be satisfied, even though the thermal conductivity of copper is relatively high. On the other hand, one may choose to suspend the sample and correspondingly use an adiabatic boundary condition. However, in this case, it is not convenient to conduct measurements under various temperature points. Therefore, the semi-infinite boundary condition is still the most practical one, as long as we ensure requirement (1) above is satisfied according to Fig. E1a.

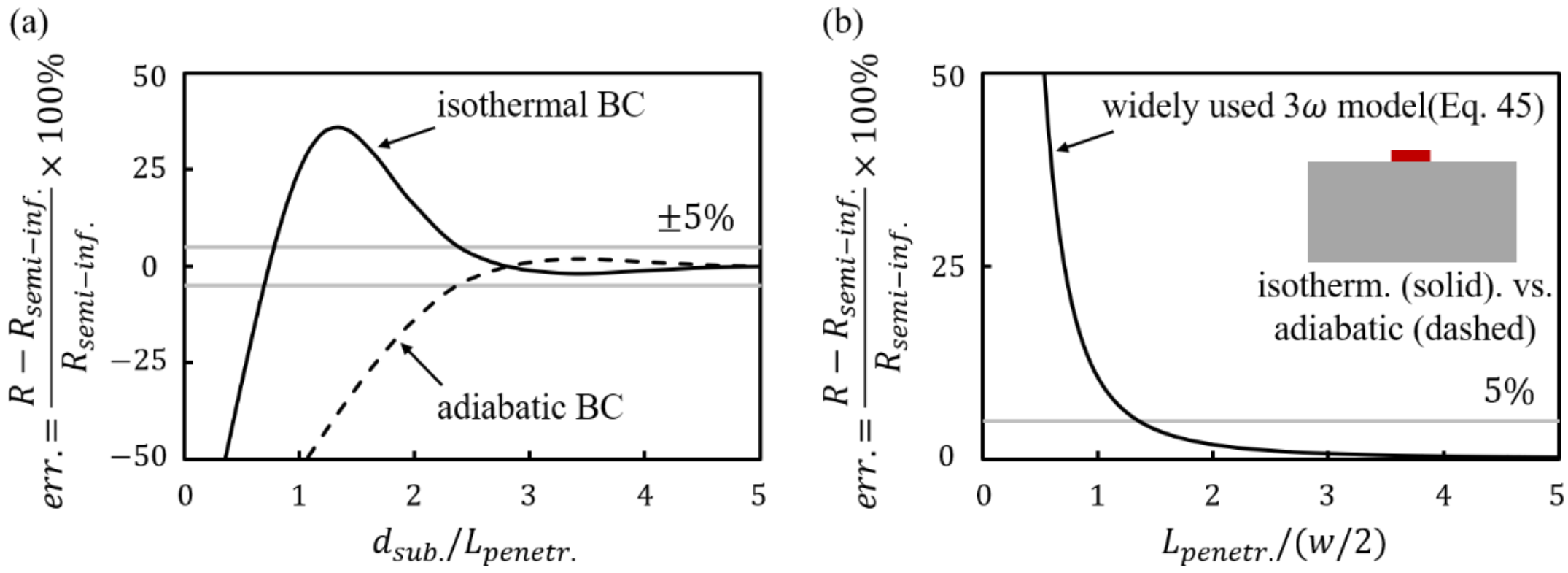

Fig. F1. Quantifying the error of the widely-used model (Eq. 44) using the general solution, Eq. 35. (a) How big of the substrate is big enough? $d_{Sub.}/L_{penetr.} > 2.40$ ensures $err. < \pm 5\%$. (b) How narrow of the heater line is narrow enough? $L_{penetr.}/(w/2) > 1.35$ ensures $err. < \pm 5\%$.

APPENDIX G: Sensitivity and uncertainty analysis

First, we conduct the sensitivity analysis to determine the important input parameters to the transfer matrix or the scattering matrix model. We define the sensitivity of the output fitting parameter, $k_{Diamond,z}$, to an input parameter of interest, $p$, as[47]

$$S_p^{k_{Diamond,z}} = \left|\frac{\partial(\ln k_{Diamond,z})}{\partial(\ln p)}\right|. \tag{G1}$$

For example, $S_p^{k_{Diamond,z}} = 2$ means a 10% change in $p$ would cause a 20% change in the "measured" $k_{Diamond,z}$. Two details are worth noting here. First, the 10% change in $p$ applies only to the transfer matrix or the scattering matrix model, but not to the FEM model. Second, the $\pm 1\%$ of "experimental noise" is not included in this sensitivity analysis.

Among the input thermal properties summarized in Table IV of the main text, Fig. G1 indicates that $k_{Diamond,z}$ is most sensitive to the substrate properties, $k_{Si}$ and $C_{Si}$, due to the fact that $k_{Diamond} > k_{Si}$ and $d_{Diamond} \ll d_{Si}$. While in the application of microprocessors, doped silicon is selected to facilitate the electrical manipulation, here in the measurement of the thermal property of the diamond film, we have the freedom to use the undoped silicon substrate that has well documented values of $k_{Si}$ and $C_{Si}$[29].

Therefore, we focus on the second most important group of input parameters, $k_{Diamond,x}$ and $R''_{c,Diamond-Si}$. Figure G1 shows that while $S_{k_{Diamond,x}}^{k_{Diamond,z}}$ depends weakly on $d_{Diamond}$, $S_{R''_{c,Diamond-Si}}^{k_{Diamond,z}}$ rapidly decreases with the increase of $d_{Diamond}$ in the thickness range of interest. The latter can be explained by a simplified resistor-series model,

$$R_{film,eff.} = \frac{d_{Diamond}}{k_{Diamond,z}(wl)} + \frac{R''_{c,Diamond-Si}}{wl}. \tag{G2}$$

$k_{Diamond,x}$ can be measured separately using other techniques, e.g. the suspended microfabricated-device method with typical measurement uncertainty of 10%[44], as mentioned in the main text and outlined in Table IV.

To obtain $R''_{c,Diamond-Si}$, we generate synthetic "experimental data" from a $0.1\mu m$-thick-diamond-film-on-Si-substrate configuration, and again invoke the 1D approximation. But this time, instead of $k_{Diamond,z}$, we extract $R''_{c,Diamond-Si}$, since in this configuration, $R''_{c,Diamond-Si}$ dominates $R_{film,eff.}$, as evident from Eq. G2. With $\pm 1\%$ "experimental noise", the "measured" $R''_{c,Diamond-Si}$ agrees with the input one to the FEM model to within 17%, as outlined in Table IV.

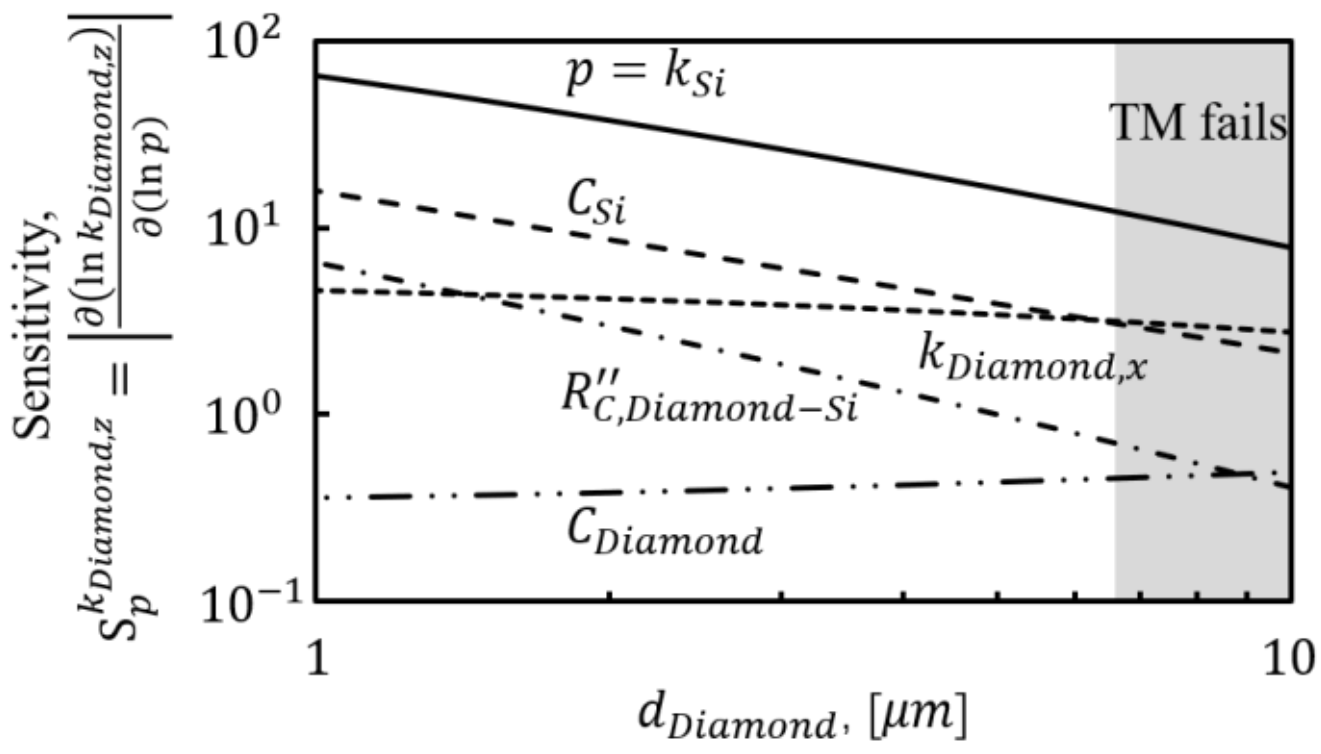

Fig. G1. Sensitivity of the "measured" $k_{Diamond,z}$ to various input thermal properties in the thickness range of the diamond film from 1 to 10 μm. While the scattering matrix model is valid in the full thickness range of interest, the transfer matrix (TM) model fails for thicker films, as indicated by the gray-shaded area. $k_{Diamond,z}$ is most sensitive to $k_{Si}$ and $C_{Si}$, but fortunately these two properties have been well studied and documented. Therefore, to reduce the uncertainty of the $k_{Diamond,z}$ "measurements" using the trasnfer matrix and the scattering matrix methods (Fig. 4), we focus on obtaining the second most important set of input parameters to the transfer matrix and the scattering matrix model: $k_{Diamond,x}$ and $R''_{c,Diamond-Si}$ (see text for details.)

Next, we conduct the uncertainty analysis. Figure 4 of the main text shows one set of synthetic "experimental data" (symbols) generated using the FEM model with $\pm 1\%$ "experimental noise" for each of the three configurations, and fits these data to either the transfer matrix (2nd column) or the scattering matrix model (3rd column) with some of the input parameters deviated from those to the FEM model, as outlined in Table IV and justified above. The only free parameter, $k_{SiO_2}$ (1st row) or $k_{Diamond,z}$ (2nd and 3rd rows), is obtained and shown in the corresponding entry of Fig. 4.

To mimic the real measurements, in which one usually conducts the experiment several times to obtain the average value and the standard deviation to represent the "true" result and the measurement uncertainty, we generate many sets of such synthetic "experiments" to obtain the statistics. For example, Fig. G2 shows the results of 100 trials of synthetic "experiments" for the scenario of the $7.5\mu m$-thick diamond film (3rd row of Fig. 4), which gives the 95% confidence interval of the vitual "experiments," 679 to 745 W/m-K.

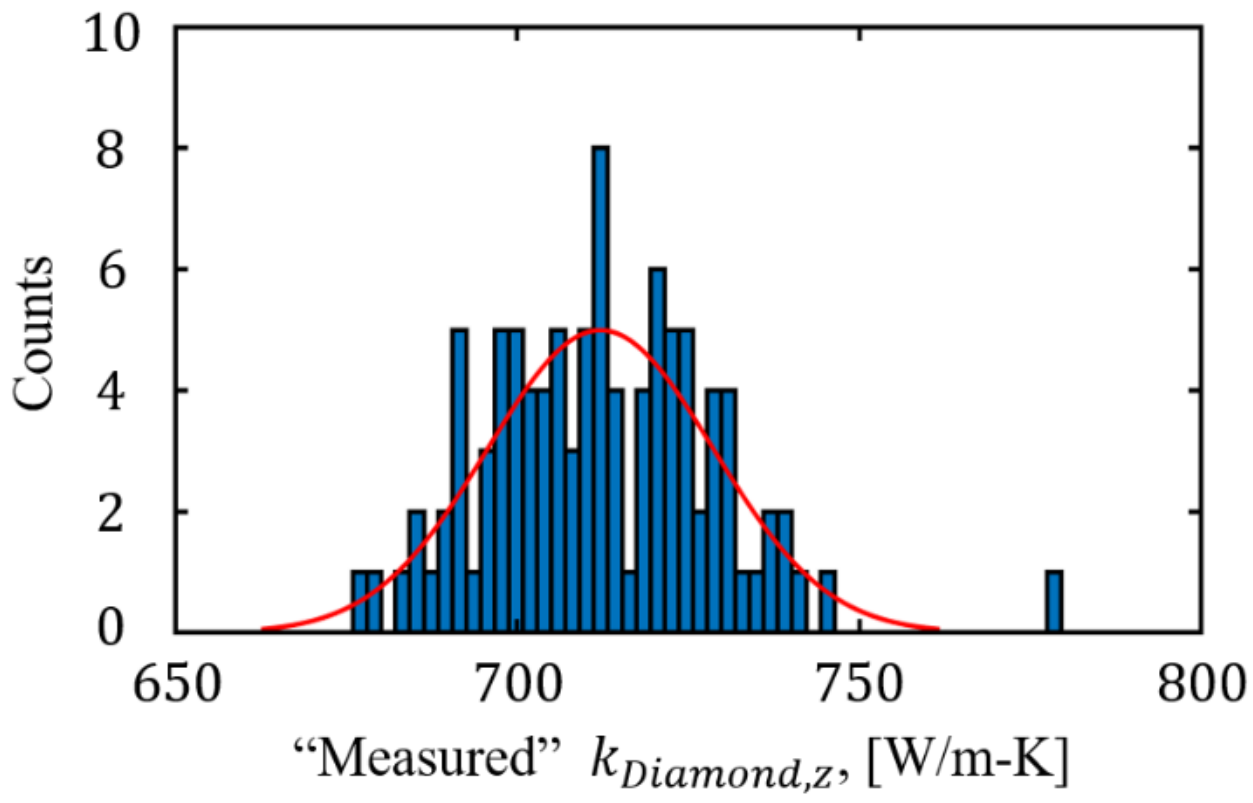

Fig. G2. Statistics of 100 trials of synthetic "experiments" on the 7.5 $\mu m$-thick-diamond-film scenario (3rd row of Fig. 4), each with $\pm 1\%$ of random "experimental noise." Some of the input parameters to the scattering matrix model are deviated from those to the FEM model, as outlined in Table IV and justified in Appendix G. The 95% confidence interval of the "measured" $k_{Diamond,z}$ is 679 to 745 W/m-K.